\documentclass{emulateapj}

\usepackage{graphicx}

\begin{document}

\title{Evidence for Infrared-Faint Radio Sources as $z>1$ Radio-Loud AGN}


\floatsep=5mm
\intextsep=5mm
\dblfloatsep=5mm
\textfloatsep=5mm
\belowcaptionskip=5mm

\author{Minh T. Huynh}
\affil{Infrared Processing and Analysis Center,, MS220-6, California Institute of Technology,
Pasadena CA 91125, USA. mhuynh@ipac.caltech.edu}

\author{Ray P. Norris}
\affil{Australia Telescope National Facility, CSIRO, Epping NSW 1710, Australia}

\author{Brian Siana}
\affil{California Institute of Technology, MS 105-24, Pasadena, CA 91125, USA}

\author{Enno Middelberg}
\affil{Astronomisches Institut, Ruhr-Universitat Bochum, Universitatsstr. 150, 44801 Bochum, Germany}

\begin{abstract}

Infrared-Faint Radio Sources (IFRSs) are a class of radio objects found in the Australia Telescope Large Area Survey (ATLAS) which have no observable mid-infrared counterpart in the {\sl Spitzer} Wide-area Infrared Extragalactic (SWIRE) survey. The extended Chandra Deep Field South now has even deeper {\sl Spitzer} imaging (3.6 to 70 $\mu$m) from a number of Legacy surveys. We report the detections of two IFRS sources in IRAC images. The non-detection of two other IFRSs allows us to constrain the source type. Detailed modeling of the SED of these objects shows that they are consistent with high redshift ($z \gtrsim 1$) AGN. 

\end{abstract}


\section{Introduction}

Infrared-Faint Radio Sources (IFRSs) were recently discovered in the Australia Telescope Large Area Survey
(ATLAS) by \cite{norris2006}. These are radio sources brighter than a few hundred $\mu$Jy at 1.4 GHz which have no
observable infrared counterpart in the {\sl Spitzer} Wide-area Infrared Extragalactic Survey (SWIRE) \citep{lonsdale2004}.  
They may be related to the optically invisible radio sources found by \cite{higdon2005, higdon2008}, which are compact radio sources with no optical counterpart to $R \sim 25.7$. 

\cite{norris2006} and \cite{middelberg2008a} have identified 53 such sources out of 2002 radio sources in the ATLAS survey. Most have flux densities of a few hundred $\mu$Jy at 1.4 GHz, but some are as bright as 20 mJy. The IFRS sources were unexpected as SWIRE was thought to be deep enough to detect all radio sources in the local universe, regardless of whether star formation or Active Galactic Nuclei (AGN) powered the radio emission. Possible explanations are that these
sources are i) extremely high-redshift ($z > 3$) radio-loud AGN, ii) very obscured radio galaxies at more moderate redshifts ($1 < z < 2$), iii) 
lobes of nearby but unidentified radio galaxies, or iv) an unknown type of object.

The nature of IFRSs has been hard to determine because they have only been detected in the radio. Spectroscopy is difficult because the hosts are optically faint and the radio positions can also have uncertainties on the order of a few arcseconds. \cite{norris2006} stacked the positions of 22 IFRSs in the {\sl Spitzer} 3.6 $\mu$m IRAC images and found no detection in the averaged image, showing that they are well below the SWIRE detection threshold. 

Recently, \cite{middelberg2008b} and \cite{norris2007} targeted 6 IFRSs with the Australian Long Baseline Array (LBA) and successfully detected two of the sources.  The \cite{norris2007} LBA detection constrained the source size to less than 0.03 arcsec, suggesting a compact radio core powered by an AGN. \cite{middelberg2008b} found the size and radio luminosity of their LBA-detected source to be consistent with a high redshift ($z > 1$) Compact Steep Spectrum Source. The VLBI detections rule out the possibility that these particular IFRSs are simply the radio lobes of unidentified radio galaxies. \cite{garn2008} stacked IFRS sources in the {\sl Spitzer} First Look Survey at infrared wavelengths as well as at 610 MHz. They find that the IFRS sources can be modelled as compact Fanaroff Riley type II (FRII) radio galaxies at high redshift ($z \sim 4$). Thus the evidence suggests that IFRSs are predominately high redshift radio-loud AGN. 

Ultra-deep {\sl Spitzer} imaging is now available over the 30 $\times$ 30 arcmin$^2$ region of the Chandra Deep Field South SWIRE field. Four IFRSs lie in this region and in this paper we report on the constraints on the nature of IFRSs derived from the new {\sl Spitzer} data.  We assume a Hubble constant of $71\,{\rm km}\,{\rm s}^{-1}{\rm Mpc}^{-1}$, and matter and cosmological constant density parameters of $\Omega_{\rm M}=0.27$ and $\Omega_{\rm \Lambda}=0.73$ in this paper.

\section{Observations, Data and Sample Selection}

The extended Chandra Deep Field South is centered at 3h32m28s, $-$27$^\circ$ 48$^\prime$ 30$^{\prime\prime}$. It overlaps the Great Observatories Origins Deep Survey (GOODS) South field, which is one of the best studied regions of the sky. The ATLAS survey \citep{norris2006} consists of deep radio observations of a 3.7 degree$^2$ field surrounding the eCDFS
which is coincident with SWIRE \citep{lonsdale2004}. The ATLAS 1.4 GHz observations reach 20 $\mu$Jy to 60 $\mu$Jy rms, with
the deepest region covering the 30 $\times$ 30 arcmin$^2$ eCDFS.  ATCA 2.4 GHz observations of the SWIRE region were obtained by the ATLAS team over the last two years and the current image reaches $\sim$ 0.1 mJy rms (Middelberg et al. in preparation). 

The SWIRE survey reaches 5$\sigma$ depths of 3.5, 7.0, 41, and 49 $\mu$Jy , respectively, in the four IRAC bands \citep{lonsdale2004}. In the MIPS bands at 24 and 70 $\mu$m the SWIRE 5$\sigma$ depths are 189 $\mu$Jy and 16 mJy. To these depths, 22 radio sources in the full SWIRE Chandra Deep Field South region were undetected, and classed as IFRS \citep{norris2006}. This paper focuses on four IFRS sources which lie in the 30 $\times$ 30 arcmin$^2$ region of the extended Chandra Deep Field South. 

The eCDFS sub-region of SWIRE has been the target of two separate {\sl Spitzer} Legacy proposals since SWIRE, and hence there are now deeper infrared data. The ``{\sl Spitzer} IRAC MUSYC Public Legacy in E-CDFSÓ (SIMPLE) Legacy project (Damen et al. in prep) provides ultra-deep imaging of eCDFS in all four IRAC bands, and the ``Far Infrared Deep Extragalactic LegacyÓ (FIDEL, PI: Dickinson) Legacy project adds ultra-deep 24 and 70 $\mu$m observations. The 5$\sigma$ depths of SIMPLE are about 0.8, 1.2, 6.3 and 6.6 $\mu$Jy in IRAC channels 1 to 4, respectively. FIDEL achieves 5$\sigma$ depths of 50 $\mu$Jy and 3 mJy at 24 and 70$\mu$m, respectively, in the areas with most coverage.
We searched for counterparts to the four IFRS sources in the new {\sl Spitzer} data. A matching radius of up to 3 arcsec was used for the IRAC and MIPS 24 $\mu$m data, and 8 arcsec for MIPS 70 $\mu$m. All four positions were examined by eye in all {\sl Spitzer} bands for any obvious failures (see Figure 1). We find two of the sources (S446 and S506) were detected in the SIMPLE ultra-deep IRAC imaging (see Figure 1) and we have improved IR constraints for the rest. Table 1 summarises the data for the four IFRSs. 

We have also searched for counterparts in existing deep optical and near-infrared (NIR) data of the eCDFS from GOODS \citep{giavalisco2004}, GEMS \citep{rix2004} and MUSYC \citep{gawiser2006}, and find an optical/NIR detection of only one IFRS source (S506). The GOODS ACS imaging reaches approximately $B_{AB} = 28.1$ , $V_{AB} = 28.9$ , $I_{AB} = 28.3$ and $z_{AB} = 27.4$. GEMS comprises two band ACS imaging and reaches depths of $V_{AB} = 28.25$ and $z_{AB} = 27.1$. The MUSYC survey has a 5$\sigma$ sensitivity of about $J_{AB} \sim 22.5$ and $K_{AB} \sim 22.0$ \citep{taylor2009}. Two IFRSs, S415 and S506, lie within the deep optical/NIR imaging area and only one, S506,  is detected as a very faint source. The lack of a detection in these optical and near-IR images for S415 strongly implies that it is at high redshift and/or is very obscured. 

A 2 Ms X-Ray exposure of GOODS South was obtained with the  Chandra X-ray Observatory \citep{giacconi2002,luo2008}. The on-axis full band sensitivity reaches 7.1 $\times$ 10$^{-17}$erg s$^{-1}$ cm$^{-2}$ at the average aim point and the minimum full band sensitivity is about 3.3 $\times$ 10$^{-16}$erg s$^{-1}$ cm$^{-2}$ over the GOODS-S region. Only one of the IFRS sources (S415) lies within the X-Ray coverage and there is no detection.

The photometry of the IFRSs was fitted with the following templates: 1) a prototypical starburst ULIRG (Arp220, \citealp{silva1998}), 2) a hot dusty AGN dominated
ULIRG (Mrk231, photometry from NED\footnote{http://nedwww.ipac.caltech.edu/. The NASA/IPAC Extragalactic Database (NED) is operated by the Jet Propulsion Laboratory, California Institute of Technology, under contract with the National Aeronautics and Space Administration.}), 3) a starforming M82-like galaxy \citep{ce01},  4) a radio-loud galaxy with an extreme radio-optical ratio (3C273, photometry from NED), and 5) an old stellar
population from \cite{maraston2005}. The old stellar templates were fitted only to the two IFRS sources with IRAC detections (S446 and S506). 
The \cite{maraston2005} templates explored have a metallicity of 0.67, assume a Salpeter IMF, and have ages of 0.3, 1 and 5 Gyr. 
The results of the Spectral Energy Distribution (SED) analysis are described in the next section.

\section{Discussion}

\subsection{S283}

This 0.287 mJy radio source lies just outside of the SIMPLE ultra-deep IRAC imaging and FIDEL 70 $\mu$m imaging. There are no SWIRE optical data either, but the shallower SWIRE IRAC 3.6  and 4.5 $\mu$m imaging shows the radio source is positioned between two IRAC sources (Figure 1), which are 3.5 and 3.7 arcsec offset, respectively. The offsets are large ($>$5 times the uncertainty of the radio position) so we do not consider either IRAC source to be a plausible counterpart. The FIDEL 24 $\mu$m imaging shows some possible flux at the source position, but it is likely that this faint flux is confusion from the two nearby IRAC sources. To verify that there is no detection of the radio source in the IRAC and MIPS 24 $\mu$m bands the two point sources were subtracted from the images and the residual image is shown in Figure \ref{residual283}. The residual flux at the radio position is $2.4 \pm 1.8$ $\mu$Jy at 3.6 $\mu$m and $13 \pm 7$ $\mu$Jy at 24 $\mu$m. This is consistent with the conservative upper limits we have assumed of 3.5 $\mu$Jy at 3.6 $\mu$m and 100 $\mu$Jy at 24 $\mu$m. 

If this source was a typical ULIRG like Arp220 or Mrk 231, even for $z > 2$, it would be detected by IRAC imaging (see Figure 3). For $z < 2$ the 24 $\mu$m flux density expected is $>>$ 100 $\mu$Jy so it would be clearly detected in MIPS. A starforming galaxy such as M82 would also have clearly been detected by {\sl Spitzer}.  While a redshift less than 1 is allowed by the 3C273 SED, that would imply a low mass object. The stellar mass estimate at $z = 1$ is $M \sim 6 \times 10^9 M_\odot$ assuming the K band luminosity from the 3C273 SED fit and a mass to light ratio of 1 (e.g. \citealp{bell2001}). This is comparable to the most massive dwarf galaxies, which are not known to be strong radio emitters or contain AGN. The IRAC limits alone suggest a stellar mass of $M \lesssim 6 \times 10^{10} M_\odot$ at $z = 1$. Radio-loud AGN with SEDs similar to 3C273 are found in massive galaxies with $M > 10^{11} M_\odot$. The IRAC non-detection therefore suggests S283 is a radio-loud AGN at  $z > 1$. 
This is supported by the galaxy's MIR-radio correlation, $q = \log(S_{24\mu {\rm m}}/S_{\rm 1.4 GHz}) < -0.46$, which is well in the range expected for radio-loud AGN (e.g. \citealp{boyle2007}).

\subsection{S415}

This IFRS has a flux density of 1.21 mJy at 1.4 GHz, but is undetected in all {\sl Spitzer} bands. Since it lies in the GOODS-S proper, where the data is most sensitive, it has by far the most extreme flux density ratios of the sources discussed here. It is as extreme, if not more so, than the optically faint  but submm bright HDF 850.1 \citep{dunlop2004, cowie2009}. This radio source has no ACS counterpart and we measured the 3$\sigma$ ACS limits at the source position to be $B_{AB} \gtrsim 28.1$. $V_{AB} \gtrsim 28.9$, $I_{AB} \gtrsim 28.3$, and $z_{AB} \gtrsim 27.4$, using a 0.6" diameter aperture. 
There is no near-infrared counterpart in the MUSYC or VLT ISAAC observations, giving 3$\sigma$ limits of $J_{AB} \gtrsim 25.5$,  $H_{AB} \gtrsim 25.8$
 and $Ks_{AB} \gtrsim 25.5$.

If this source has an SED similar to M82, Arp220 or Mrk 231 it would be detected in all {\sl Spitzer} bands, even 70 $\mu$m, for $z < 4$. If the radio-loud 3C273 SED is adopted, this source must lie at $z \gtrsim 3$ for it to remain undetected in the IRAC and ACS bands (assuming no obscuration, see Figure 4). At redshift $z = 1$, the IRAC 3.6 and 4.5 $\mu$m limits are a factor of 2 to 3 below the 3C273 SED, which implies an extinction of 0.8 to 1.1 mags would be required for this source to be undetected in the IRAC channels. This level of extinction is seen in some extreme ULIRGs (e.g. \citealp{genzel1998, murphy2001}), but this source is not luminous in the infrared. This source therefore lies at redshift $z >> 1$. Even at these redshifts obscuration of a few magnitudes is required for a 3C273 object  to be undetectable in the optical/NIR bands (see Figure 4). At $z \gtrsim 1$ the radio luminosity of S415 is $P_{\rm 1.4 GHz} > 5 \times 10^{24}$ W Hz$^{-1}$. The {\sl Spitzer} non-detection of S415 therefore suggests that this source is a distant ($z >> 1$) obscured radio-loud AGN. 

The X-Ray non-detection implies that this source has an X-Ray luminosity $L_{0.5-8keV	} \lesssim 2 \times 10^{42}$ erg s$^{-1}$ at $z = 2$ and $L_{0.5-8keV} \lesssim 6 \times 10^{42}$ erg s$^{-1}$ at $z = 3$, where the \cite{luo2008} limit assumes an X-Ray power law of $\Gamma = 1.4$. This suggests that, for these redshifts, either the source has a very high column density of neutral hydrogen obscuring the X-Ray emission, which is consistent with the optical/NIR non-detection, or the source is intrinsically X-Ray faint for an AGN. 

A preliminary analysis of ATCA 2.4 GHz data (Middelberg et al. in preparation) finds S415 has a significant detection ($> 7\sigma$) of 0.67 mJy. This implies a steep radio spectral slope of $\alpha = -1.1$ {\footnote{$S \propto \nu^\alpha$}, which can not be produced by star forming processes. S415 has a slope similar to some ultra steep spectrum sources (\citealp{rottgering1997, debreuck2004}), which have been linked to massive high-redshift ($z > 2$) radio galaxies.

\subsection{S446}

IFRS S446 is a 0.338 mJy radio source which lies just outside the GEMS ACS coverage and MUSYC near-infrared imaging. There is a faint source in the IRAC 3.6  and 4.5 $\mu$m channels just 2.2 arcsec north of the radio position (see Figure 1) and we assume this is the counterpart to S446. The probability that one or more IRAC sources lies randomly within a distance $\theta$ of a radio source is \( P = 1 - \exp(-\pi n \theta^2)\), for an IRAC source density $n$ (often called the $P$-statistic; e.g. \citealp{downes1986}). The P-statistic is only a rough estimate because it does not does not take into account the individual positional uncertainties and assumes a random distribution of the background population, whereas astronomical sources are clustered. Nevertheless, for the SIMPLE IRAC source density of 116700/deg$^2$ the P-statistic suggests that the chances of an IRAC source lying closer than 2.2 arcsec is 13\%, so the counterpart is reasonably reliable. The source is 6.6 $\pm$ 0.3 $\mu$Jy and 5.7 $\pm$ 0.5 $\mu$Jy at 3.6 and 4.5 $\mu$m, respectively, and not detected in the other IRAC bands. There is no detection in the 70 and 24 $\mu$m images, but limits are hard to quantify because the source falls in between 2 brighter sources. We assume the noise here is twice that of the local noise in the MIPS imaging for the purposes of the SED fitting.

Similar to the other IFRS sources, the non-detection in the two longer wavelength IRAC bands, and MIPS 24 and 70 $\mu$m bands, rules out M82, Arp220 and Mrk 231 SEDs for this source, at a wide range of redshift ($z < 6$). A 3C273 SED would not produce the MIR emission detected in the IRAC bands (see Figure 5), so a possible explanation for this source is a radio-loud AGN which dominates the radio emission but with a stellar component which is seen in the MIR. 

The  \cite{maraston2005} old stellar population fit to the IRAC data constrains the redshift of this galaxy to 1 -- 1.5, and the best fit is a 1 Gyr old model at $z = 1.5$. At this redshift the radio luminosity of S446 is $P_{\rm 1.4 GHz} = 3.7 \times 10^{24}$ W Hz$^{-1}$, which would place it at the low power end of radio-loud AGN. 

Using \cite{maraston2005} SEDs with a different metallicity, including a different library of SEDs (e.g. \citealp{bruzualcharlot2003}), or adding reddening as a free parameter would give a different best-fit redshift, but this level of complexity can not be explored with only two datapoints.  Lastly we note that the IRAC detections are unlikely to be caused by the hot dust component of an AGN torus. The IRAC detections imply the MIR peak is shorter than 3.6 $\mu$m, so the dust temperatures would be greater than $\sim$1600 K at $z  = 1$ and greater than $\sim$2400 K at $z = 2$, using Wien's law. Silicate grains sublimate at about 1000 K, while graphite grains sublimate around 1500 K, so these temperatures are too high for an AGN torus. Instead, AGN torus models typically show a cooler MIR emission peak of 7 to 10 $\mu$m (e.g. \citealp{schartmann2005}), which is longward of IRAC channels 1 and 2. 

\subsection{S506}

This source is similar to S446. This 0.170 mJy radio source has a GEMS ACS V and z band detection 1.4" south of the radio position that is 26.27 and 25.62 AB magnitudes, respectively (see Figure 1). The likely counterpart in the IRAC 3.6 and 4.5 $\mu$m channels is less than 1" away from the ACS source, roughly 2.3" south of the radio position. This ACS counterpart, while faint, is a 7.0 and 5.1 sigma detection in the V and z bands, respectively. 

 The $P$-statistic suggests the chance of this IRAC source being a random coincidence is about 14\%. The source is 5.5 $\pm$ 0.3 $\mu$Jy and 5.5 $\pm$ 0.4 $\mu$Jy at 3.6 and 4.5 $\mu$m, respectively, and not detected in the other IRAC bands. There is no MIPS 70 or 24 $\mu$m detection in the FIDEL images of this source. The FIDEL image shows a possible 24 $\mu$m excess but this has a flux density of only 7.3 $\pm$ 4.2 $\mu$Jy and it  is not coincident with the IRAC source. There is also no detection in the MUSYC NIR imaging ($J_{AB} \gtrsim 22.5$ and $K_{AB} \gtrsim 22.0$).  Again, M82, Arp220 and Mrk 231 type SEDs are ruled out for this source by the non-detection in the MIPS 24 and 70 $\mu$m bands.  The MIR peak is between 3.6 and 4.5 $\mu$m and if hot dust is responsible for this then Wien's law implies hot dust temperatures of $\sim$1400 K at redshift 1 and $\sim$2200 K at redshift 2. Similar to S446, this rules out the IRAC detection of a hot AGN tori. The most likely explanation for this source is a radio-loud AGN at ($z >2$ ) with a stellar component dominating the IRAC channels (Figure 6). 

Using the IRAC data alone, we find the best fit \cite{maraston2005} old stellar population model to the IRAC data is 1 Gyr old and places this galaxy at $z = 2.5$.  
The 5 Gyr model has a redder SED that would be detected by IRAC 5.8 $\mu$m imaging at z $<$ 2, and the 1 Gyr model becomes a bad fit for $z > 2.6$. However the ACS detection suggests a blue excess from star formation activity. The best fit stellar population model to both ACS and IRAC detections is a 0.3 Gyr \cite{maraston2005} model at $z = 2.0$.  As for S446, the small number of datapoints does not warrant exploring all the free parameters in the redshift fitting, such as metallicity and reddening. However, SEDs younger than 0.3 Gyr were also explored for S506 and these young SEDs have a blue excess inconsistent with the faintness of S506 in the optical bands. At the best fit redshift of $z = 2.0$, S506 has a radio luminosity $P_{\rm 1.4 GHz} = 3.6 \times 10^{24}$ W Hz$^{-1}$, which, similar to S446, would place it at the low power end of radio-loud AGN. 

The ACS counterpart is made up of two clumps (see Figure 1), one of which is extended and has faint emission over 0.3 arcsec, which is 2.5 kpc at redshift $z = 2$. The dual nature of S506 could be explained by one component having an AGN while one (or both) has some recent star formation.   

\subsection{IR-radio Correlation}

The IR-radio correlation (e.g. \citealp{yun2001}) is an indicator of the dominant emission mechanism in a galaxy because both IR and radio emission are thought to be strongly linked to star formation. Any deviation from the correlation is a sign of an AGN. Galaxies with excess radio emission probably contain a radio-loud AGN, while IR-excess sources are likely to be radio-quiet AGN with hot dust dominating in the MIR (for $z > 1$).  The observed MIPS to radio flux density ratio limits are shown in Figure 7. Some of the xFLS galaxies (Frayer et al. 2006) have ratios consistent with the IFRSs but all are easily detected at optical and infrared wavelengths. The $S_{24\mu{\rm m}}/S_{\rm1.4 GHz}$ limits for all 4 IFRSs show that they have excess radio emission for $z \lesssim 2$. The limits from 70 $\mu$m are not as stringent as that from  24 $\mu$m, but they do show that the IFRSs have as much excess radio emission as the AGN-dominated ULIRG Mrk231. The extreme source S415 has a ratio that is consistent with a radio-loud AGN at $z \gtrsim 4$. The IFRS IR-radio ratios are consistent with a sample of high redshift radio galaxies (HzRGs, \citealp{seymour07}), which host luminous radio-loud AGN. 

\subsection{Could IFRSs be galactic objects?}

We now consider whether IFRSs may be galactic objects such as pulsars or radio stars. The majority of known pulsars are young objects with a low spin rate and these are distributed across the Galactic disc, as expected for objects that have originated fairly recently from massive stellar supernovae. They are rare objects at high galactic latitude. For example, the Parkes Multibeam Pulsar Survey, with a 1.4 GHz flux density limit of 0.15 mJy for long period pulsars, finds a density of about 1/deg$^2$ for $|b| < 1$ \citep{camilo2000}. The density of these pulsars drops to $<$0.25/deg$^2$ by $|b| = 4$ deg. Applying the P-statistic from Section 3.3, we find the probability of one or more pulsars within 15 arcmin (half the eCDFS size) of a radio position is less than 5\%, assuming a pulsar densityof  $<$0.25/deg$^2$. The eCDFS has a galactic latitude of about -55 degrees, so the space density of pulsars in the eCDFS field is much less than that. Therefore the 0.25 degree eCDFS field is unlikely to contain a pulsar. 

Shorter period pulsars, so-called millisecond pulsars, are thought to have been spun up to high rotation rates as a member of a low-mass X-ray binary (LMXB) system. These objects are usually found in globular clusters (e.g \citealp{manchester2001}). The eCDFS is not a globular cluster field and so is unlikely to contain a LMXB. 

The optical non-detection of the radio sources gives us another clue about whether the sources are pulsars. 
Neutron stars have been detected in the optical, and we can take two examples: the Crab pulsar and the Vela pulsar. The Crab pulsar's optical counterpart is known as Baade's star and it is relatively bright in the optical, with V = 16.6. The Crab pulsar lies at 2 kpc \citep{manchester2005}, so at 10 kpc it would have V = 20.1 and  V = 25.1 at 100 kpc. Thus, if these IFRSs are as bright as the Crab pulsar in the optical they would be easily detectable in the optical imaging. The Vela pulsar is much fainter however, V = 23.6 \citep{mignani2001}, and lies at a distance of 294 pc \citep{caraveo2001}. An object such as this would be too faint for eCDFS HST detection if it lies further than about 2.3 kpc. 

X-Ray emission from the neutron star surface or from the the pulsar wind nebula maybe detectable if the IFRS is a pulsar. The one source with X-Ray data, S415, has an X-Ray luminosity limit that is 100 times less than some of the faintest X-Ray emitting pulsars known (e.g. \citealp{kargaltsev2009}). X-Ray emission from a neutron star may not be orientated the same way as the radio emission, and the X-Ray-radio observations were not simultaneous, so this is not conclusive evidence that S415 is not a pulsar. It is however consistent with the idea that S415 is an extragalactic source. 

Could IFRSs be main sequence stars? Ultracool dwarf stars, a class thought to be radio active, were observed at 4.8 GHz, but only one was detected in a survey of eight \citep{antonova2008}. This star lies at a distance of 12.2 pc and has a 4.8 GHz flux density of 0.286 mJy. \cite{berger2006} observed 90 M and brown dwarf stars and found 8.5 GHz flux densities well below 1 mJy, although flares can increase the flux density by at least a factor of a few. The stars in this sample are all closer than about 13 pc and have J and K magnitudes brighter than 18 mag.  The typical M-dwarf has an absolute magnitude of 8 - 17 $M_V$ (e.g. \citealp{kaler1997}), so at a distance of 1 kpc (10 kpc) M-dwarfs have an observed magnitude of V = 18 - 27 (23.1 - 32.1). Therefore the weakest of these would be undetected in the optical images at large distances, but the radio studies suggest M-dwarf stars close enough to be detected in the radio would be seen in the optical images. 

\subsection{Comparison to other galaxy populations}

We can compare the space density of our sample of IFRSs with that of other high redshift $z \sim 2$ samples. For example, much work has been done on the $BzK$ selection technique \citep{daddi2004} to select star-forming and passive galaxies at $z \sim 2$. The space density of $BzK$s is $\sim$ 1 per arcmin$^2$ \citep{daddi2007}, more than 200 times greater than that of the IFRSs in this work ($\sim$16 per deg$^2$). 

Another high redshift sample is  Dust-Obscured Galaxies (DOGs), which are selected using a combination of red colors ($R - [24] > 14$, in Vega magnitudes) and bright flux densities in infrared ($S_{24} > 0.3$ mJy) \citep{dey2008}. These criteria proved remarkably efficient for selecting z $\sim$ 2 galaxies.  DOGs contribute $\sim$ 26\%  of the total IR luminosity density at $z = 2$ and 60\% of the total from  ULIRGs. They have a space density of $\sim$ 0.1 per arcmin$^2$ or $2.82 \pm 0.05 \times 10^{-5}$ h$^3$ Mpc$^{-3}$, similar to submillimeter galaxies and about 20 times more numerous than the IFRSs.

The paucity of IFRSs is not a surprise as they were selected to have extreme radio to infrared ratios, and as such are rare compared to other populations selected in the optical and infrared. How do AGN samples selected in the radio compare? By matching Faint Images of the Radio Sky at Twenty-cm (FIRST) 1.4 GHz radio sources with the Sloan Digital Sky Survey (SDSS), \cite{ivezic2002} find about 3 radio selected AGN per deg$^2$ with  $S_{1.4} > 1$ mJy and $i < 21$. The four IFRSs in this study, likely radio-loud AGN at $z > 1$, have a source density  $\sim$16 per deg$^2$. This suggests that IFRSs are a population of radio-loud AGN which have been unstudied by previous radio work. 

S446 and S506 lie at redshifts 1 to 2 and have radio powers ($P_{\rm 1.4 GHz}$) of $\sim$  10$^{24}$ W Hz$^{-1}$. The local luminosity function of radio AGN \citep{best2005, sadler2002} suggests that there are 40 to 60 AGN  per deg$^2$ with radio powers of 10$^{24}$ - 10$^{25}$ W Hz$^{-1}$ in the volume between redshifts 1 to 2, assuming no evolution. If luminosity evolution (e.g. \citealp{donoso2009, sadler2007}) is applied the number of AGN with radio powers of 10$^{24}$ -10$^{ 25}$  W Hz$^{-1}$ increases to 110 to 140 per deg$^2$. Therefore, the IFRSs presented in this work make up only a small portion ($\lesssim$ 1\%) of the total AGN population with similar radio powers at that epoch. The remainder are presumably bright enough to be detected in the {\sl Spitzer} bands. 

The IFRSs can also be compared to high redshift radio galaxies(HzRGs). A representative sample of 69 HzRGs at $z > 1$ were observed with {\sl Spitzer} \citep{seymour07} and these massive galaxies ($10^{11}$ - $10^{11.5}$ M$_{\odot}$) have the mid-IR luminosities of LIRGs or ULIRGs. While HzRGs have much greater radio flux densities than IFRSs, they have extreme MIR-radio ratios consistent with the available data on IFRSs (Figure 7). Hence, IFRSs maybe the lower luminosity analogs of HzRGs. There are only a few hundred known HzRGs across the full sky, so they are even rarer than IFRSs.

\section{Summary}

We have searched for infrared counterparts to four Infrared Faint Radio Sources in the extended Chandra Deep Field South field using recently available ultra-deep {\sl Spitzer} observations. No IFRS is detected in ultra-deep 24 $\mu$m imaging, implying that the sources do not follow the IR-radio correlation and star formation can not produce all the radio emission observed in these objects. We find IRAC detections for two of the sources, and the non-detections of the other two provides constraints on the source SEDs. Typical ULIRG SEDs, such as Mrk231 and Arp220, and $L_*$ galaxies are ruled out by the {\sl Spitzer} data. The most likely explanation for these sources is that they are radio-loud AGN, with radio-to-optical ratios similar to 3C273, at redshifts $z \gtrsim 1$. The most extreme source (S415) lies at  $z >> 1$ and requires several magnitudes of obscuration in the optical/NIR to remain undetected by deep imaging. It is very unlikely that the IFRSs are Galactic sources, but current data cannot conclusively rule this out in the case of the undetected sources.

For the two sources with IRAC detections we find the SED can be described by a radio-loud 3C273-like SED combined with a stellar population.  The stellar population SED fits to the IRAC MIR data, combined with the available optical limits, suggest that the two detected IFRS sources have redshifts of $z \sim 1.5$ and $z \sim 2.0$.  For one source, S506, the ACS detection suggests the galaxy may have a stellar component that is only 0.3 Gyr in age. These two sources are similar to the IRAC detected OIRSs which are posited to be ``red and dead" radio galaxies at $z > 1$ \citep{higdon2008}.

The evidence is mounting that a significant proportion, if not all, of the IFRS sources are radio-loud AGN at high redshift ($z \gtrsim 1$), and not merely lobes of an unidentified radio galaxy. The source density in eCDFS is $\sim$16 per deg$^2$ for $S_{1.4} > 0.1$ mJy. \cite{garn2008} find a source density of  $\sim$3.5 per deg$^2$ for $S_{1.4} > 0.5$ mJy. While these sources are rare, they point to a population of AGN at high redshift that has been undiscovered until recently.

\acknowledgements 

This work is based in part on observations made with the {\sl Spitzer Space Telescope}, which is operated by
the Jet Propulsion Laboratory, California Institute of Technology under a
contract with NASA.  Support for this work was provided by NASA through an award issued by JPL/Caltech.  
This research has made use of the NASA/IPAC Extragalactic Database (NED) which is operated by the Jet Propulsion Laboratory, California Institute of Technology, under contract with the National Aeronautics and Space Administration.

\clearpage

\bibliographystyle{aj}
\bibliography{paper_refs}

\begin{figure*}[hbt]

{\footnotesize 
\begin{tabbing}
\hspace*{1.9cm} \= \hspace*{3.4cm} \= \hspace*{3.4cm} \= \hspace*{3.5cm}   \= \hspace*{5.3cm} \kill\> 
3.6 $\mu$m \> 4.5 $\mu$m \>   24 $\mu$m \> ACS V\\
\end{tabbing}
}

\vspace{-4mm}

\begin{minipage}{0.08\textwidth}
{\footnotesize S283}
\end{minipage} 
\hspace{-7mm}
\begin{minipage}{0.9\textwidth}
\includegraphics[width=3.4cm]{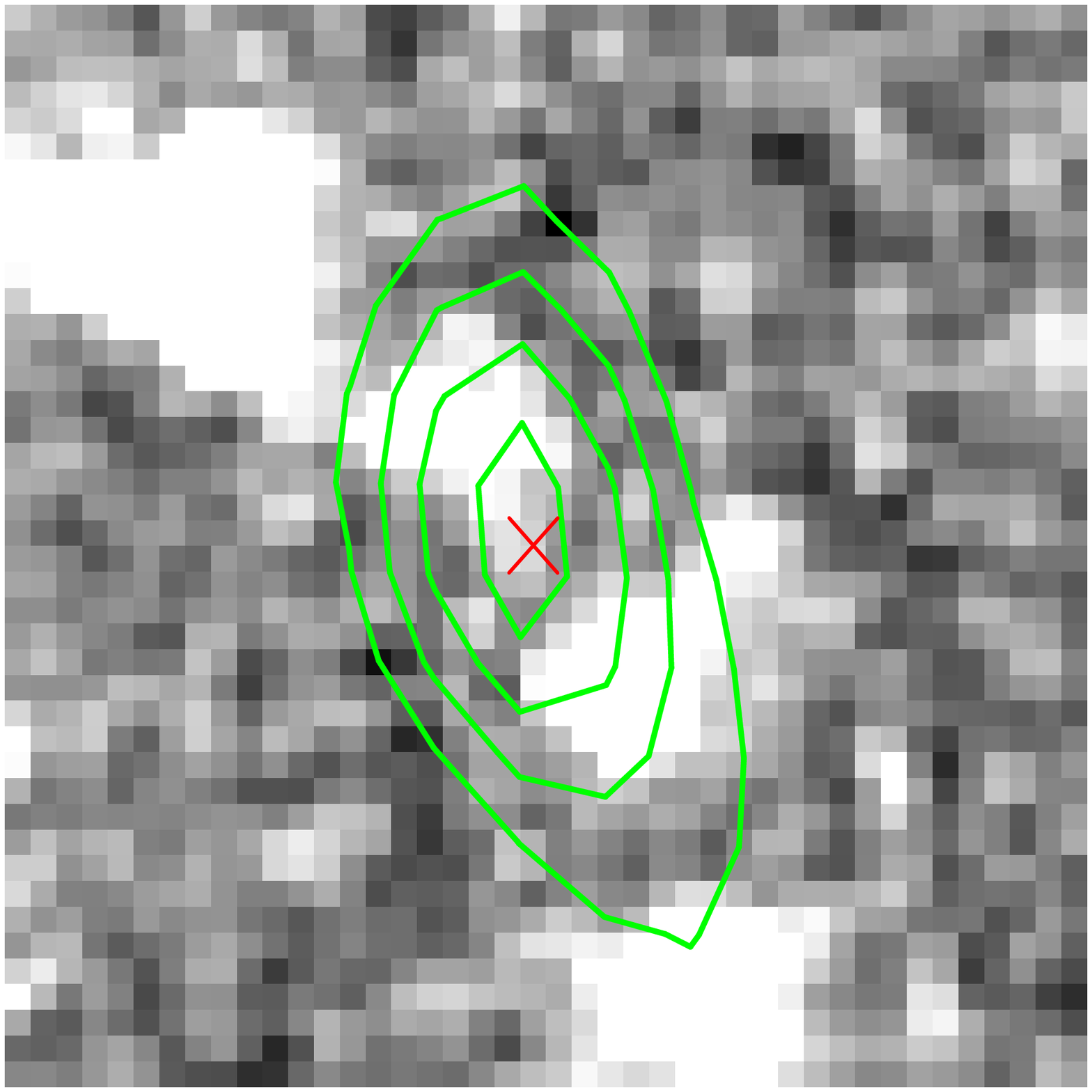}
\includegraphics[width=3.4cm]{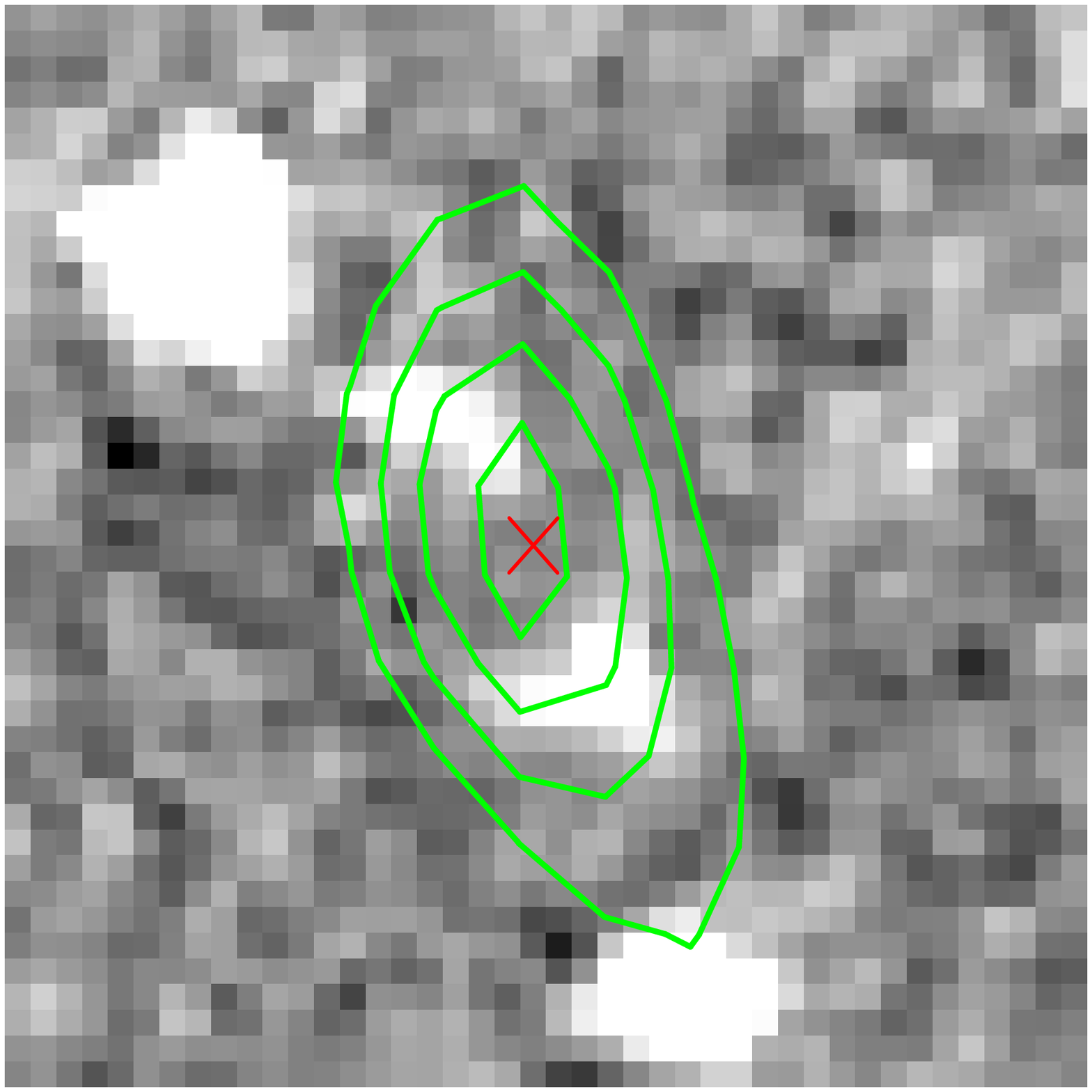}
\includegraphics[width=3.4cm]{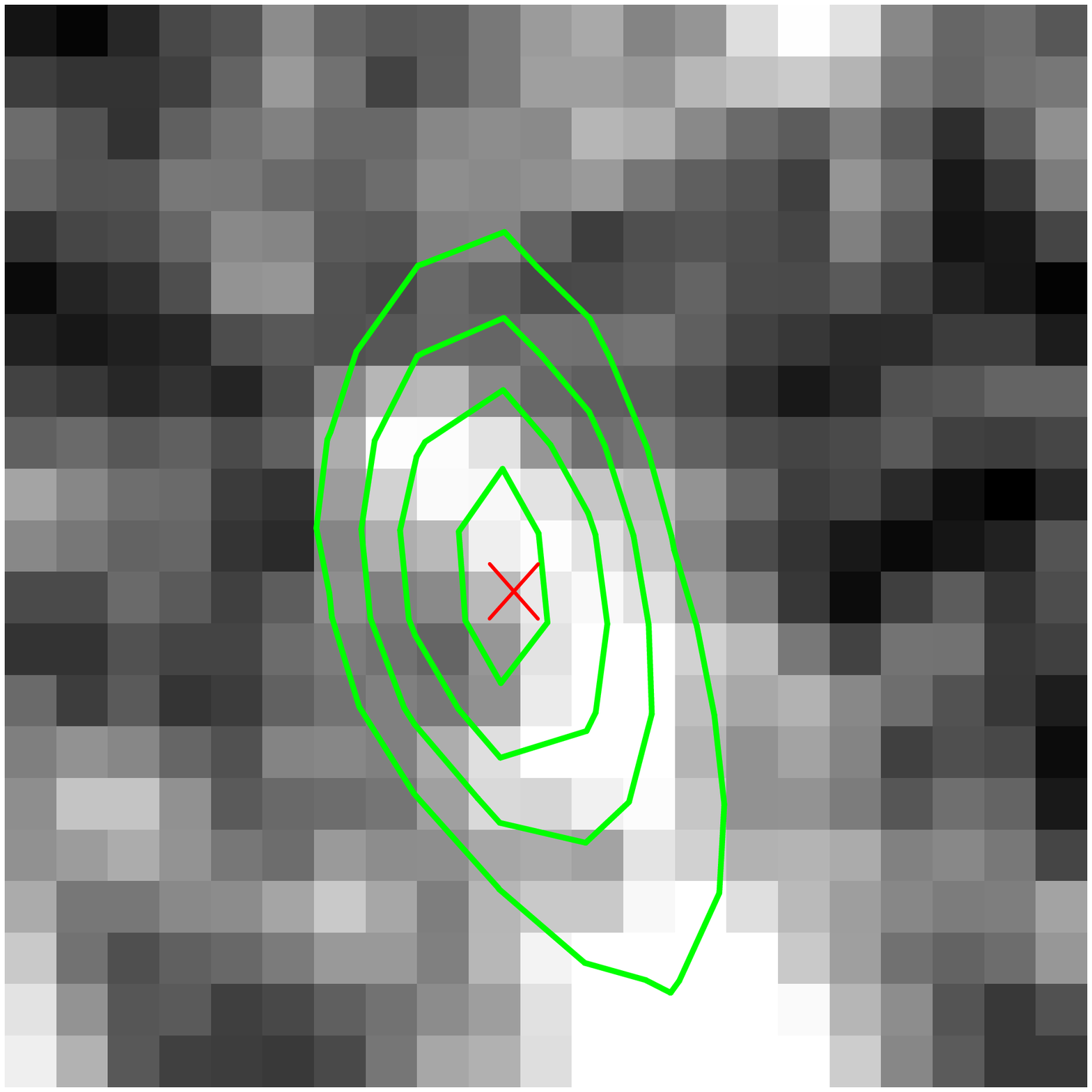}
\end{minipage}

\vspace{1mm}

\begin{minipage}{0.08\textwidth}
{\footnotesize S415}
\end{minipage} 
\hspace{-7mm}
\begin{minipage}{0.9\textwidth}
\includegraphics[width=3.4cm]{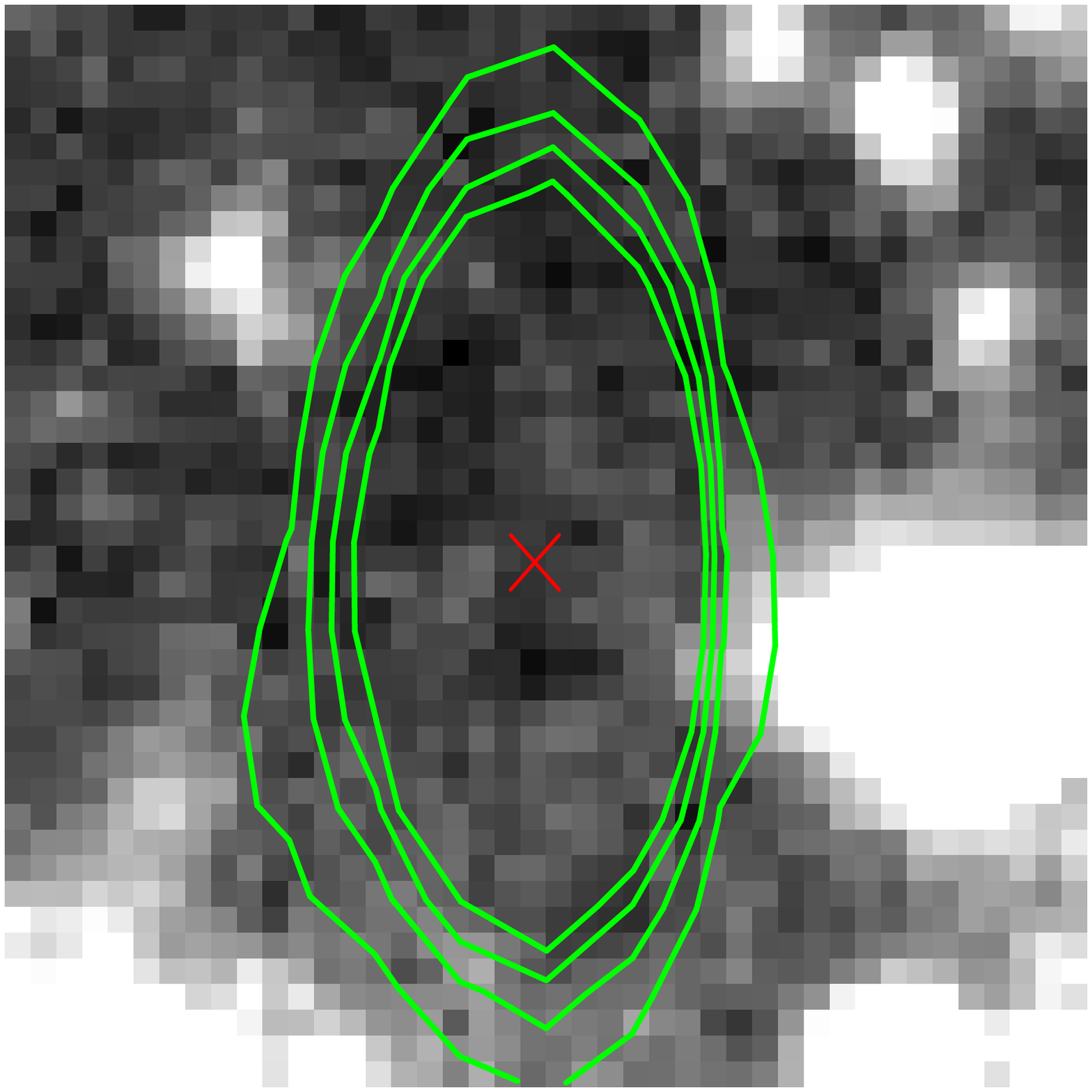}
\includegraphics[width=3.4cm]{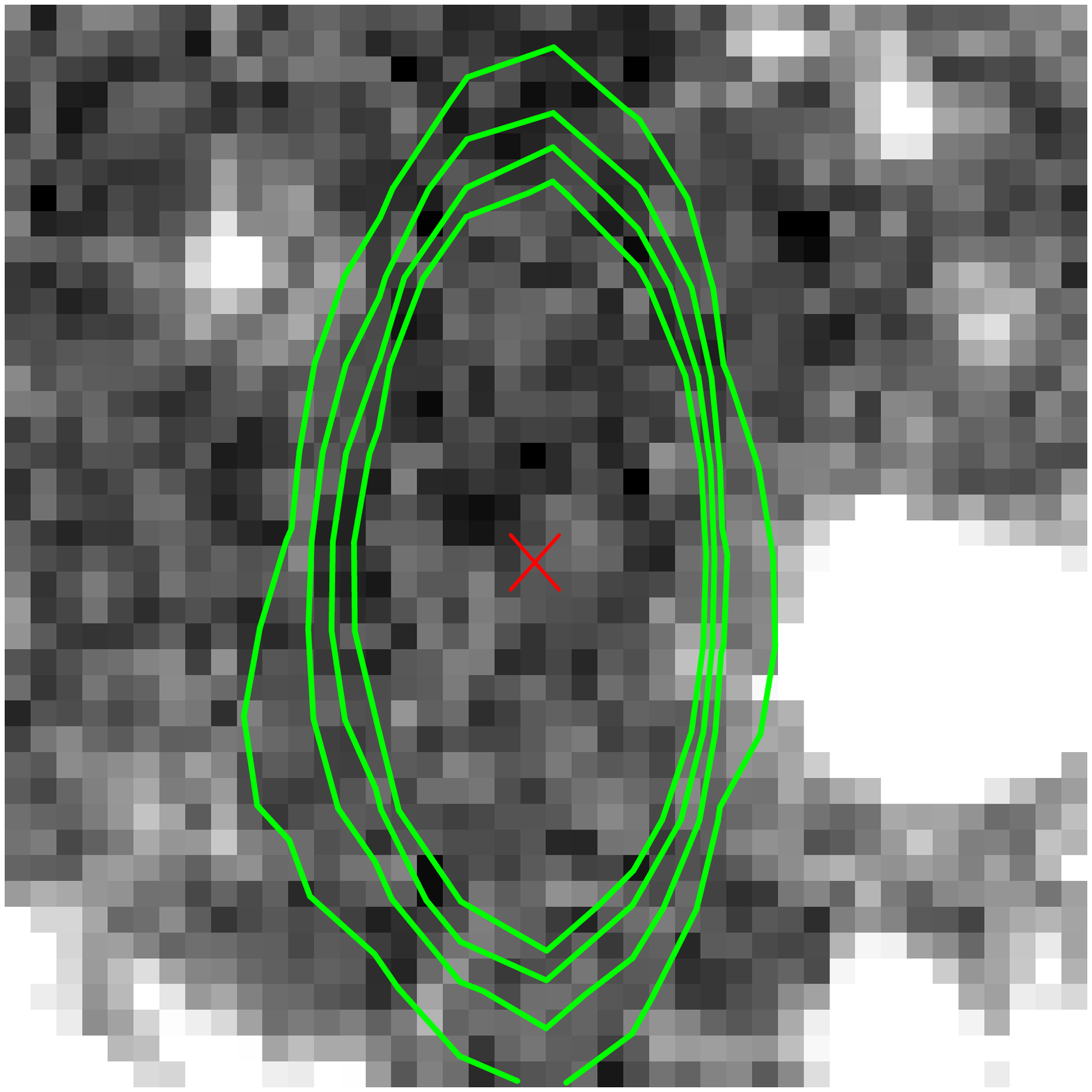}
\includegraphics[width=3.4cm]{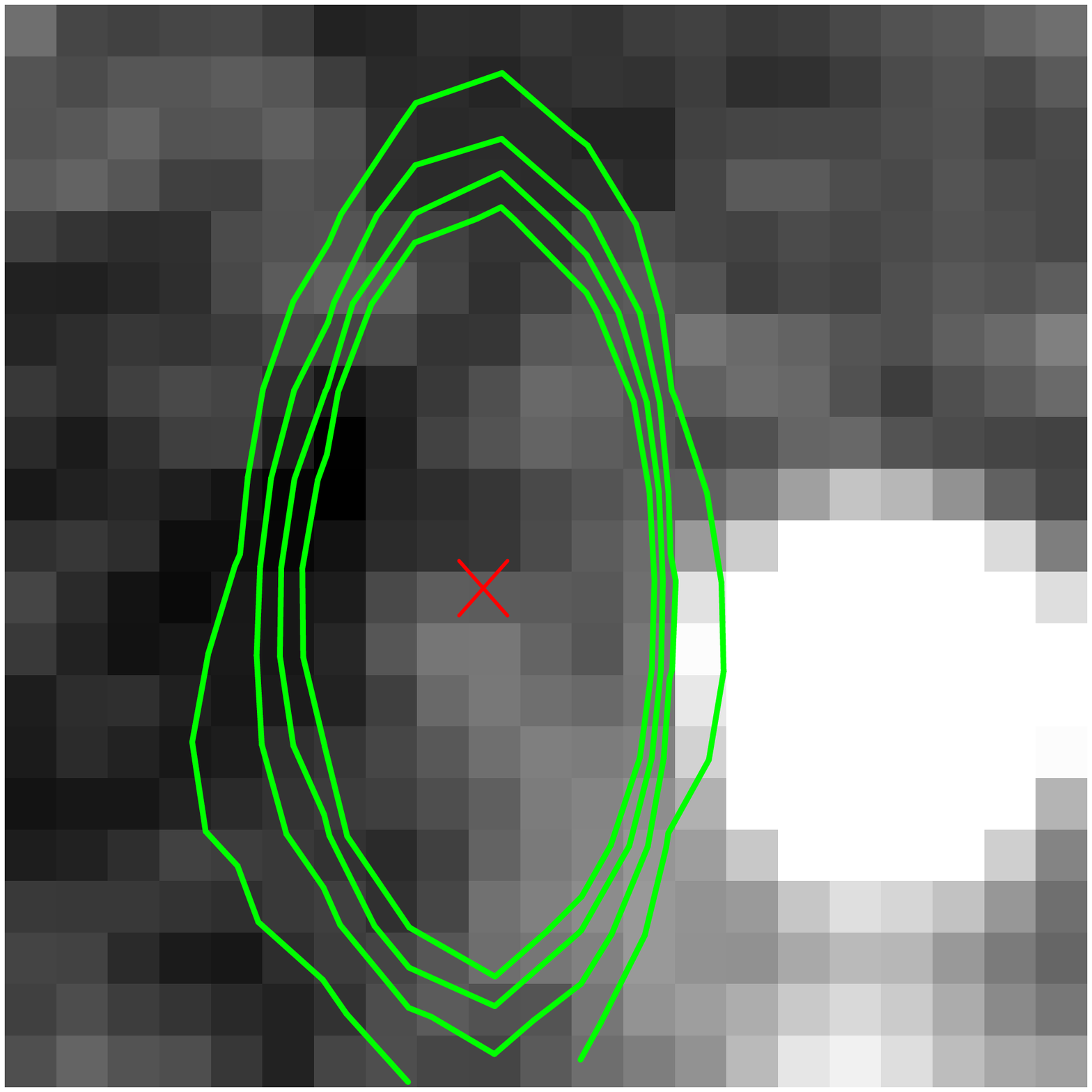}
\hspace{3mm}
\includegraphics[width=3.2cm]{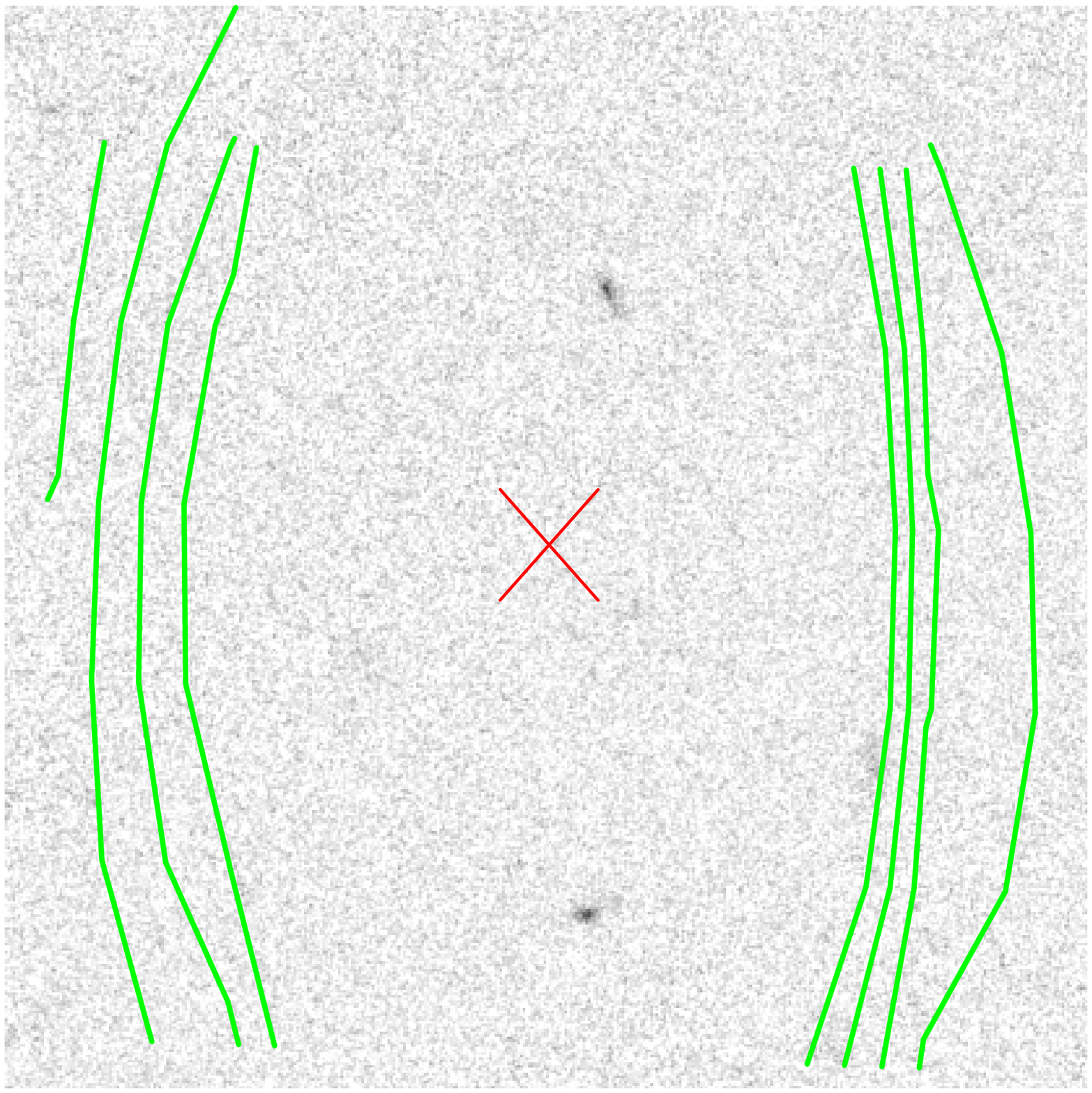}
\end{minipage}

\vspace{1mm}

\begin{minipage}{0.08\textwidth}
{\footnotesize S446}
\end{minipage} 
\hspace{-7mm}
\begin{minipage}{0.9\textwidth}
\includegraphics[width=3.4cm]{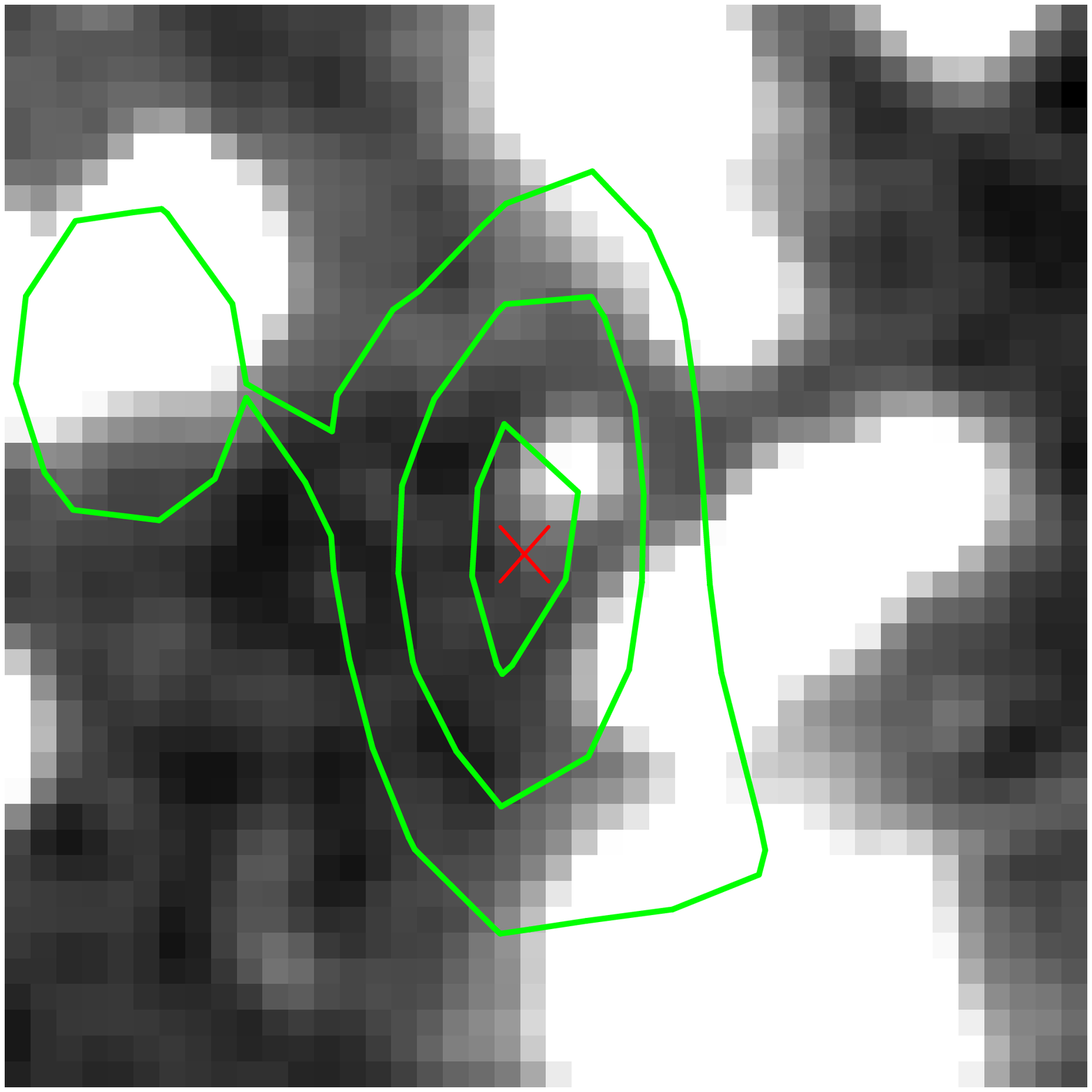}
\includegraphics[width=3.4cm]{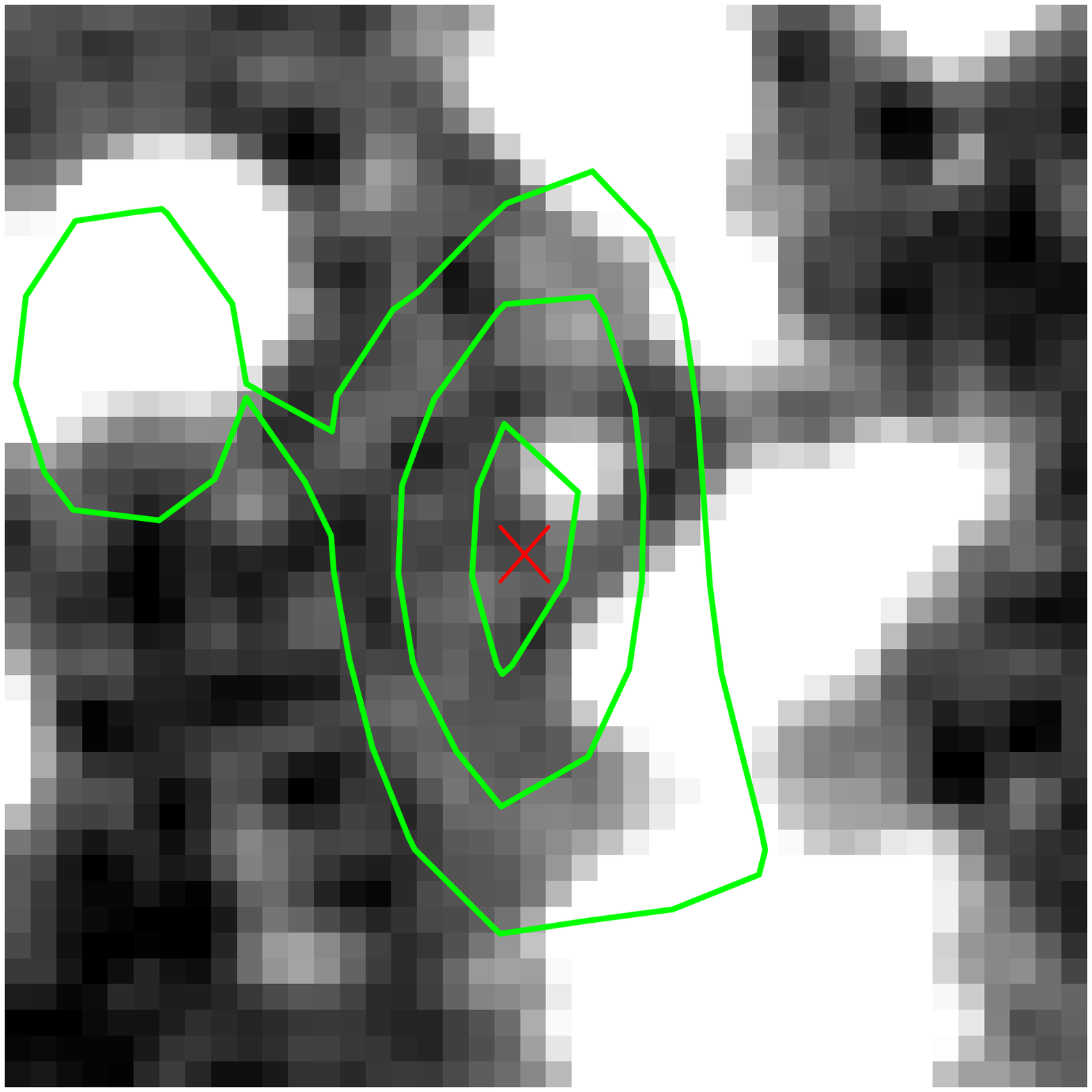}
\includegraphics[width=3.4cm]{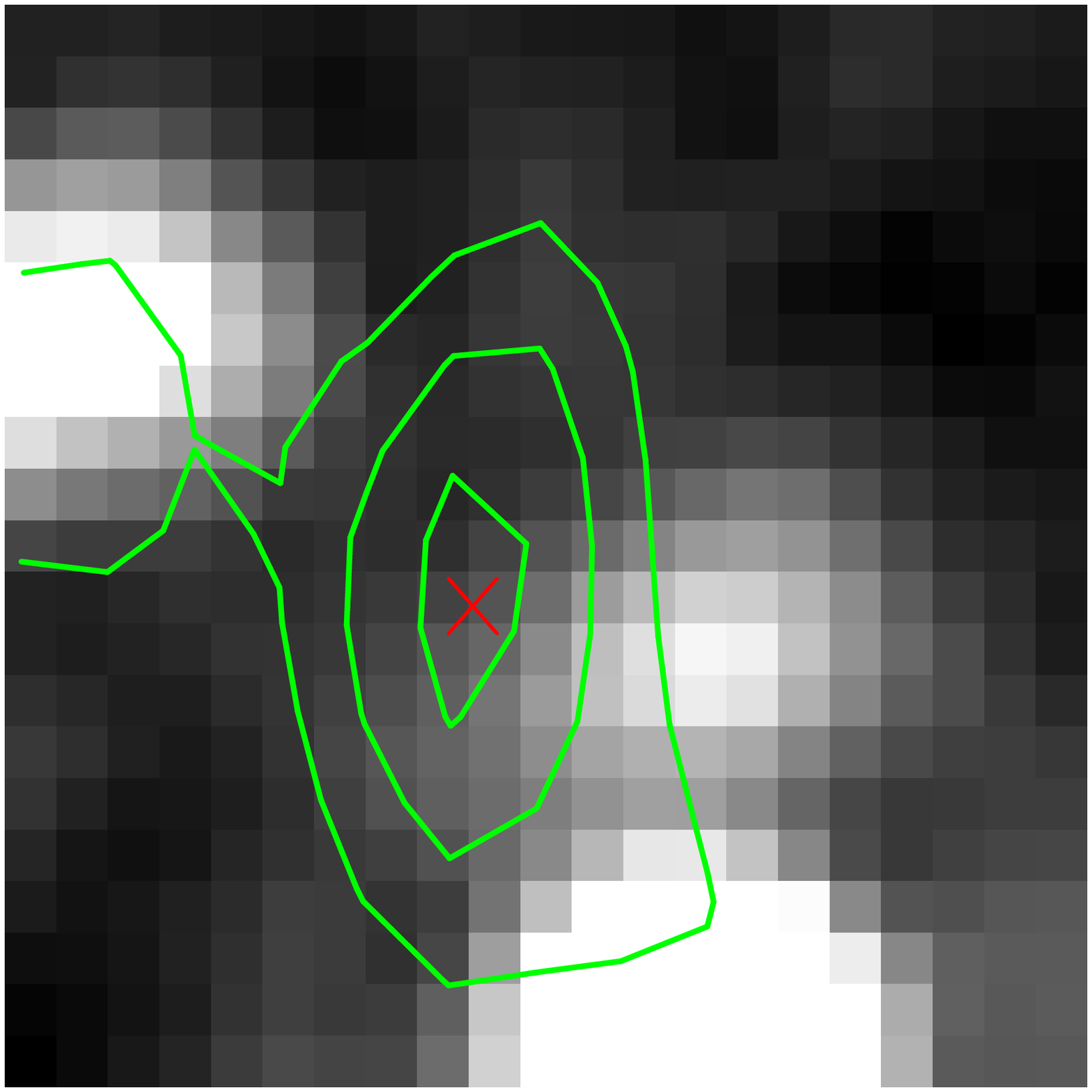}
\end{minipage}

\vspace{1mm}

\begin{minipage}{0.08\textwidth}
{\footnotesize S506}
\end{minipage} 
\hspace{-7mm}
\begin{minipage}{0.9\textwidth}
\includegraphics[width=3.4cm]{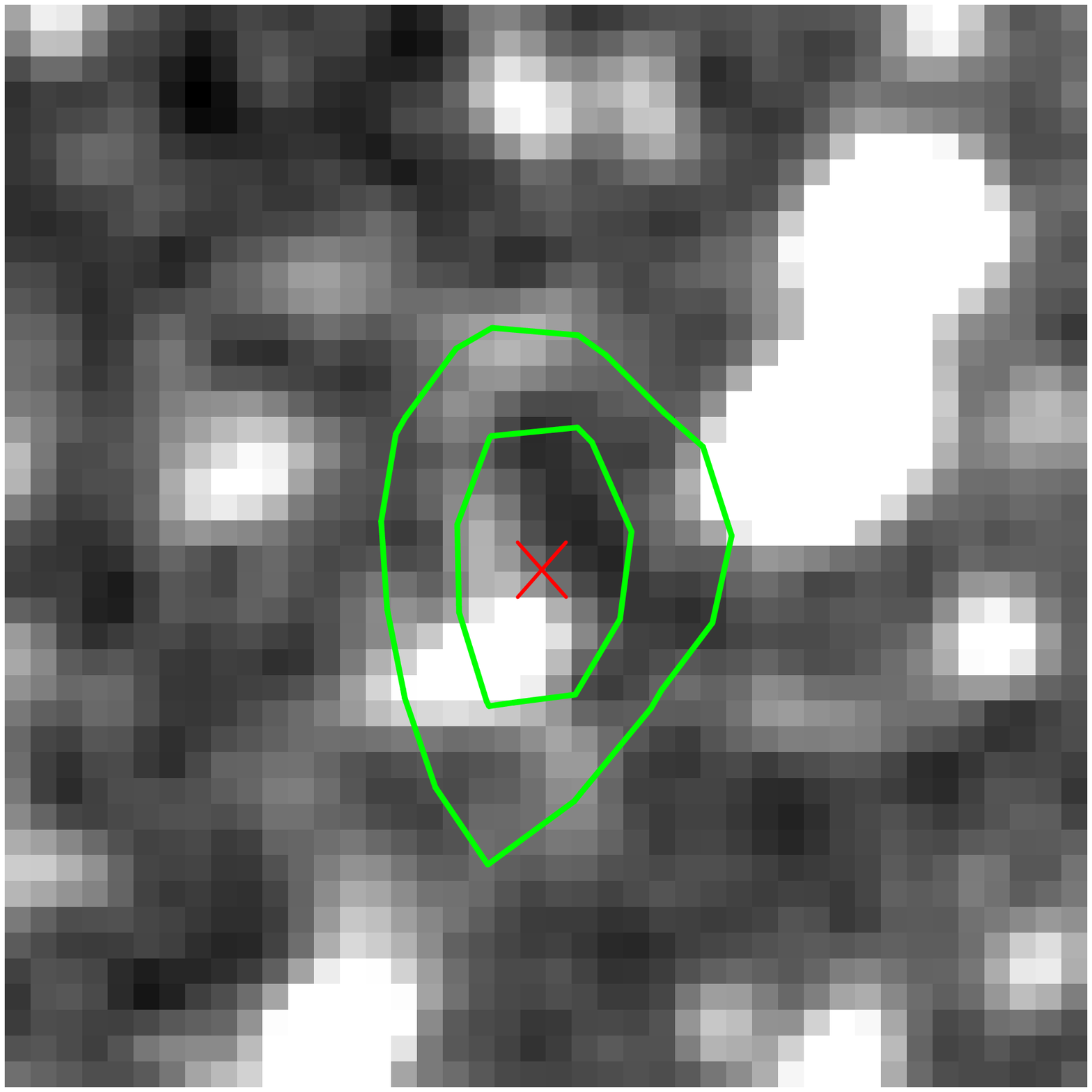}
\includegraphics[width=3.4cm]{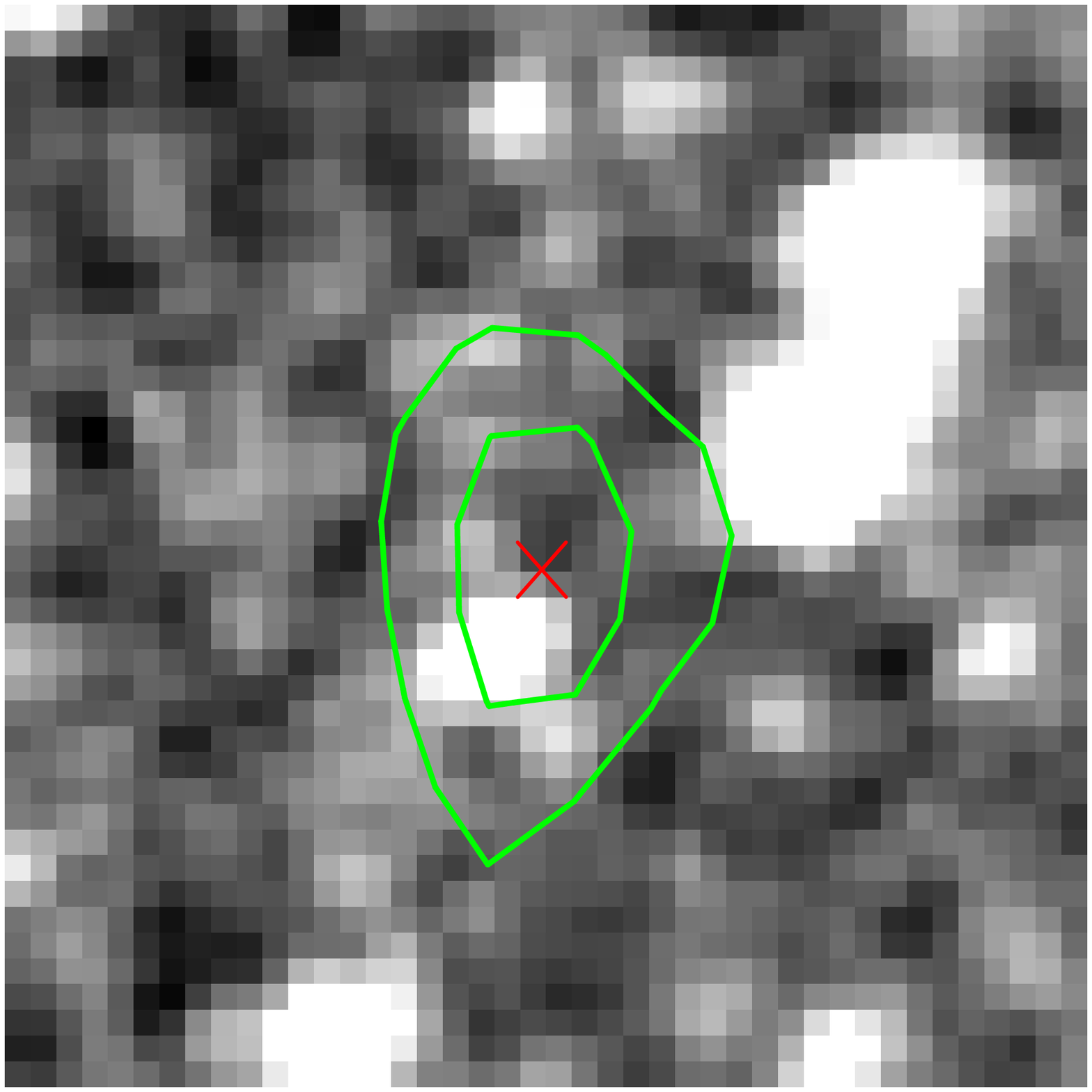}
\includegraphics[width=3.4cm]{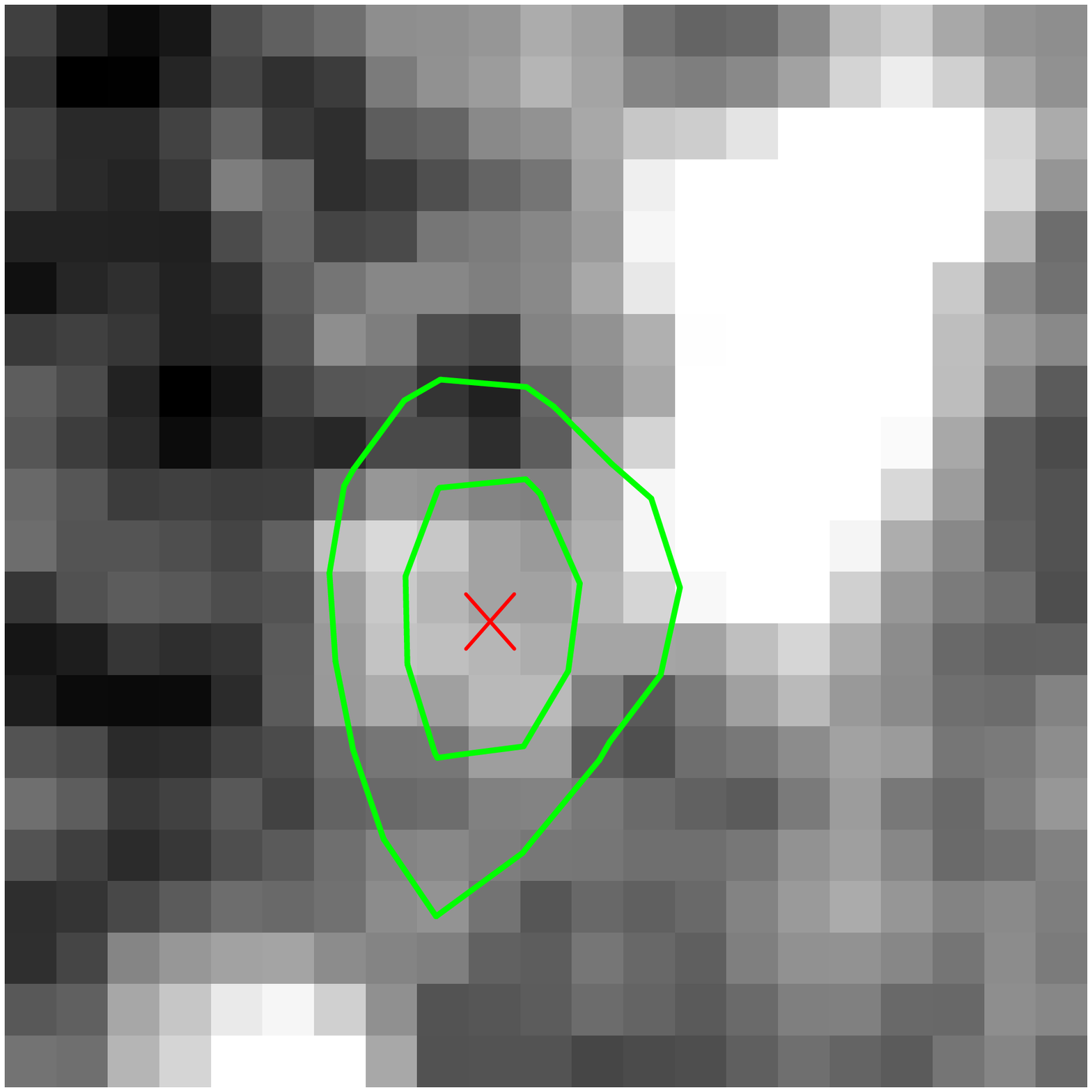}
\hspace{3mm}
\includegraphics[width=3.2cm]{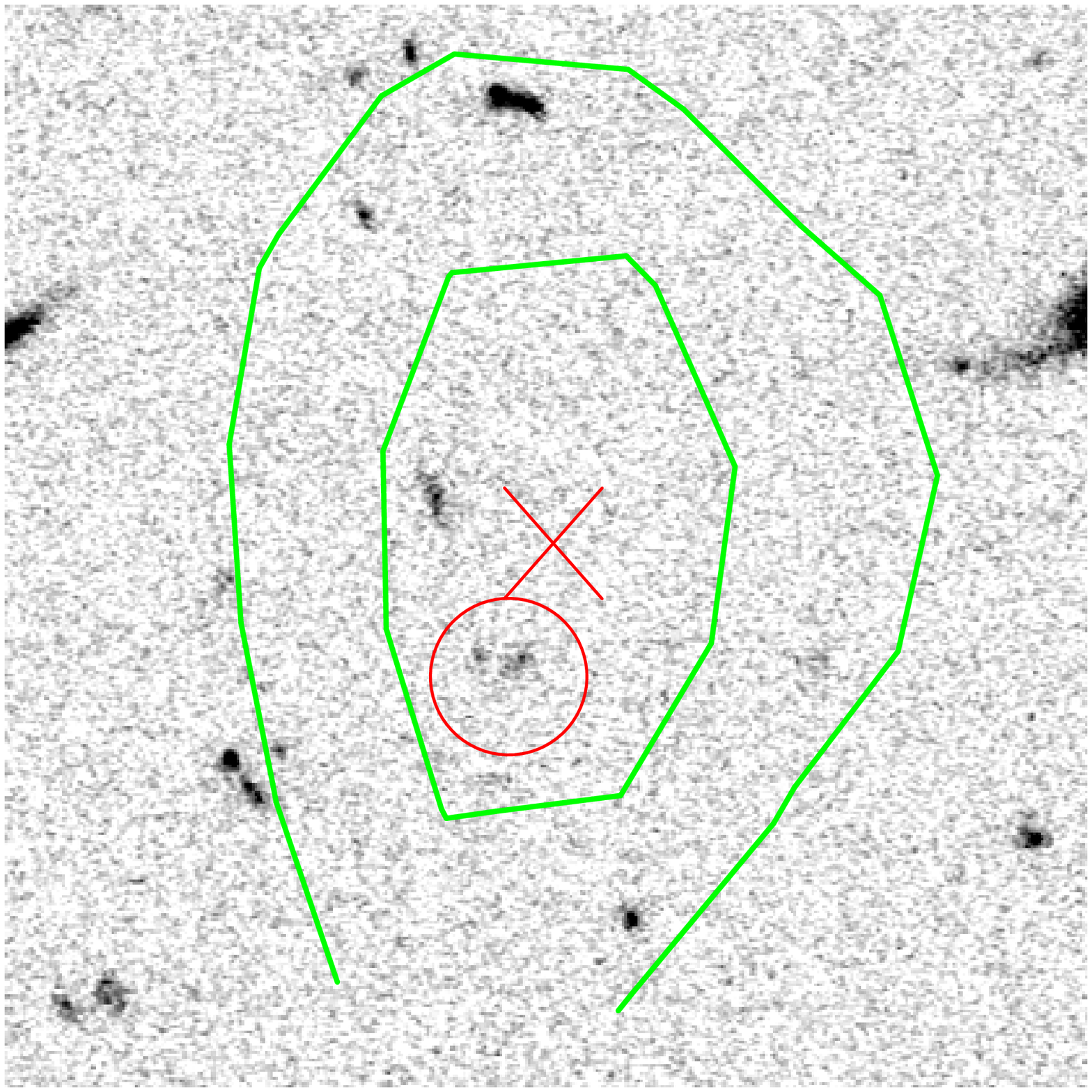}
\end{minipage}

\caption{Grey scale {\sl Spitzer} and ACS images of S283, S415, S446 and S506. From left to right, images are IRAC 3.6, 4.5 $\mu$m, MIPS 24 $\mu$m, and ACS V band. The ACS grey scale is inverted for clarity. The green contours are the radio 1.4 GHz image with levels set at 50, 100, 150, 200 $\mu$Jy. The red cross marks the radio position. The {\sl Spitzer} images are 25 $\times$ 25 arcsec and the ACS image is 12.5 $\times$ 12.5 arcsec. The IRAC counterpart to S446 can be seen 2.2 arcsec north of the radio position. The red circle in the S506 ACS image mark the very faint optical counterpart, which is consistent with the proposed IRAC counterpart 2 arcsec south of the radio position.}
\end{figure*}

\begin{figure}[hbt]
\includegraphics[width=3.3cm]{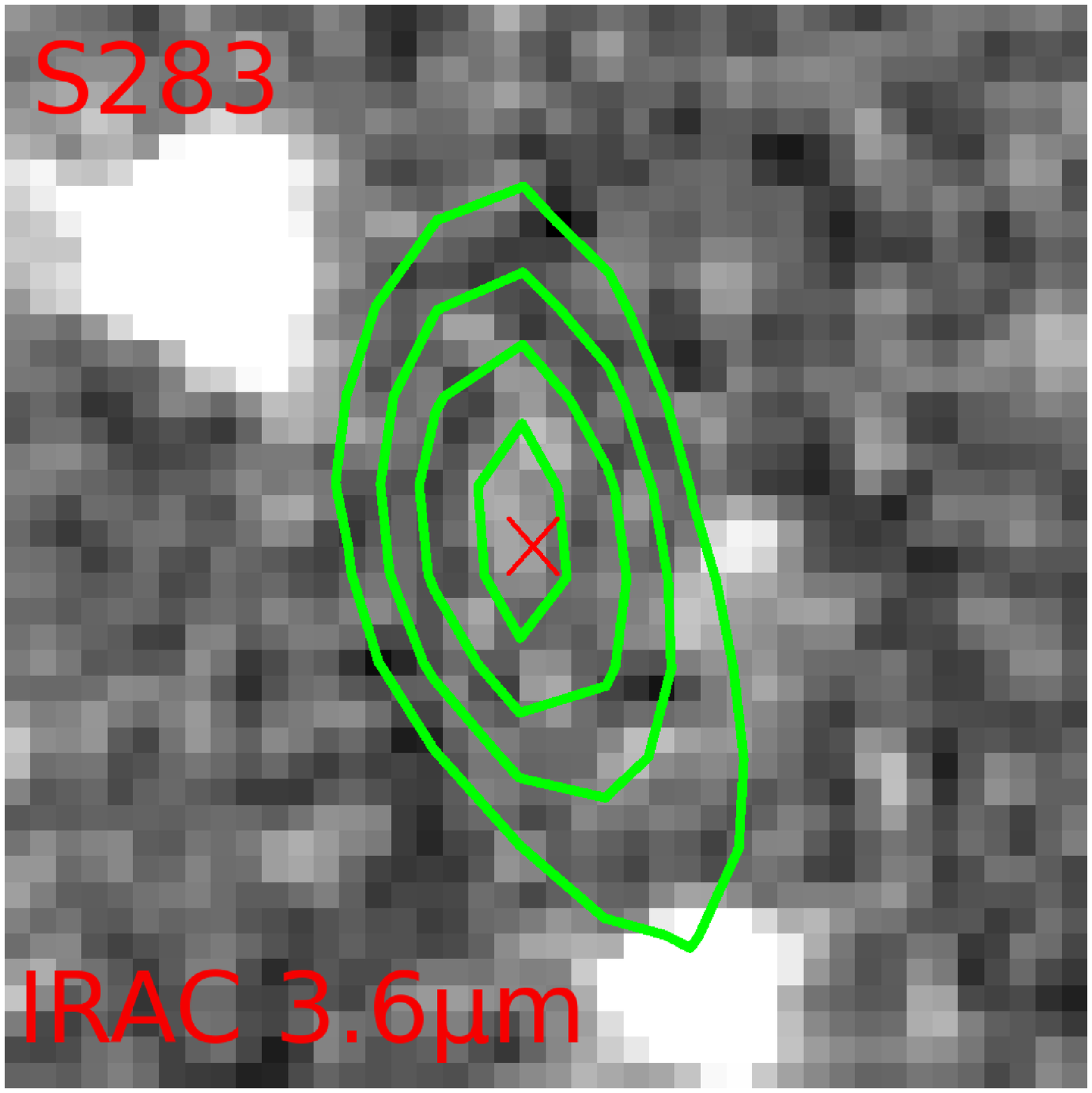}\vspace{2mm}
\includegraphics[width=3.3cm]{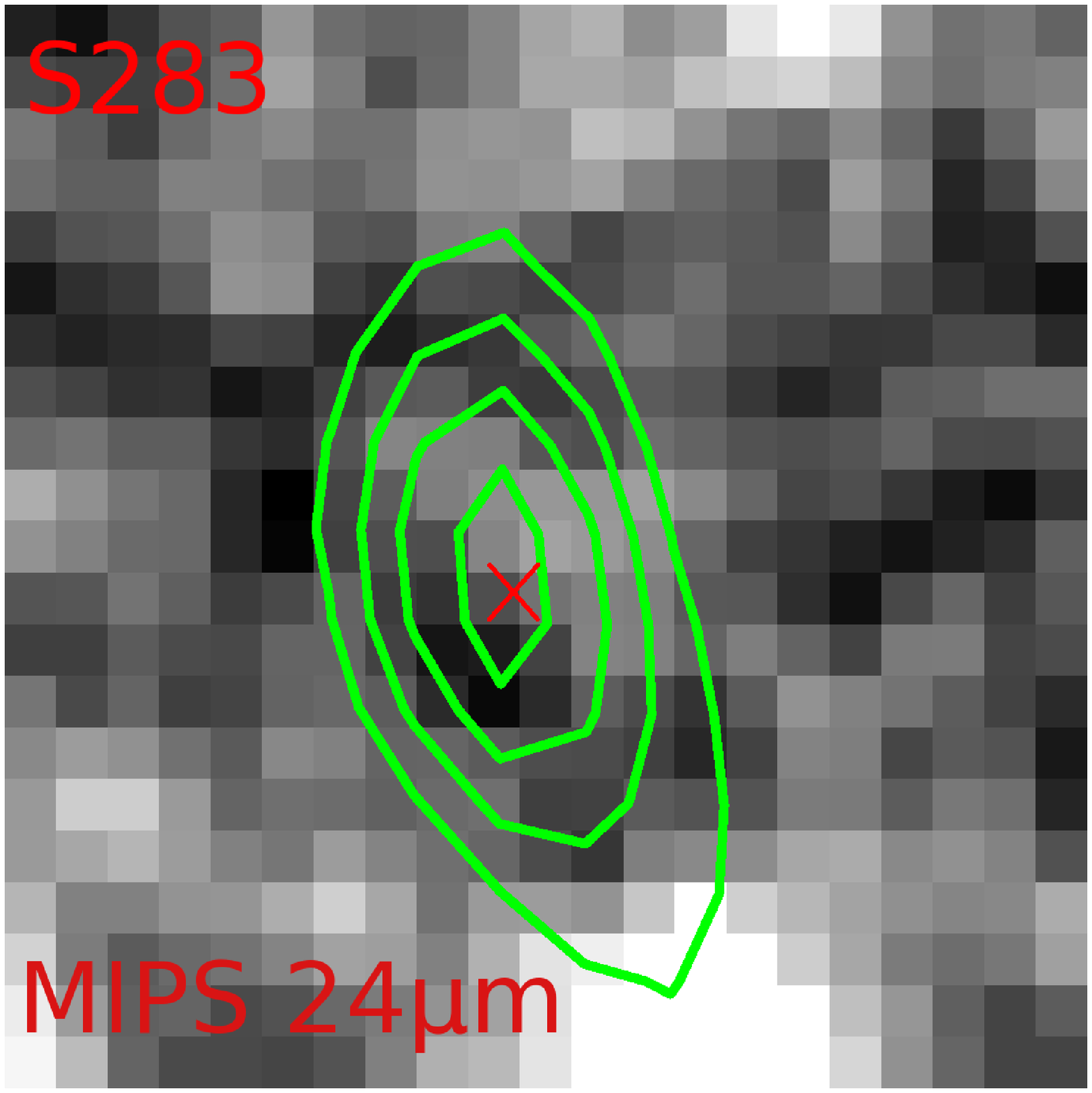}
\caption{The residual image for S283 with the two point sources near the radio center removed, for IRAC 3.6 $\mu$m (left) and MIPS 24 $\mu$m (right). As for Figure 1, The green contours are the radio 1.4 GHz image with levels set at 50, 100, 150, 200 $\mu$Jy  and the red cross marks the radio position. There is very little IRAC or MIPS flux at the radio position.}
\label{residual283}
\end{figure}

\begin{figure}[hbt]
\includegraphics[width=10.5cm, viewport=25 5 495 345]{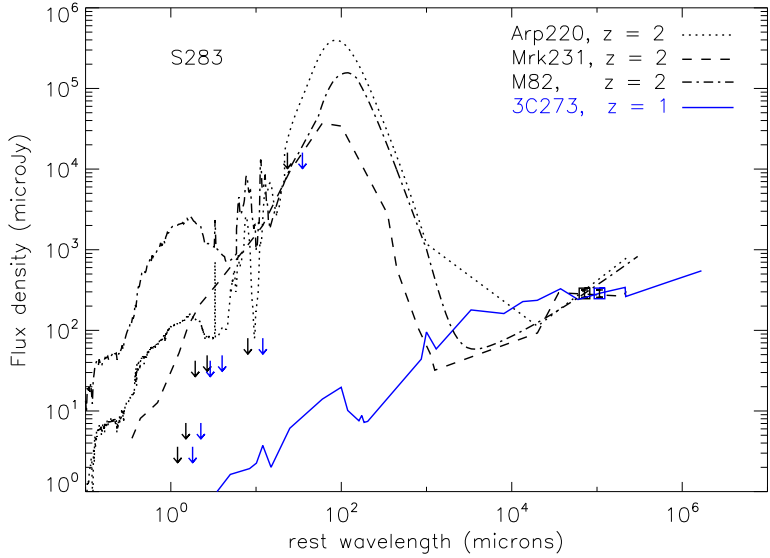}
\caption{The SED of S283, fit by Arp220, Mrk231, M82 and 3C273. Points in black assume the galaxy is at $z = 2$, and
blue are the data for $z = 1$. This galaxy is not well fit by M82, Arp220 or Mrk231, as it would be detected by IRAC and
24$\mu$m imaging. The SED of radio-loud galaxy 3C273 at $z > 1$ is consistent with the observations. The SEDs are scaled up in luminosity by 90, 12 and 2000 times for Arp220, Mrk231 and M82 respectively, and 3C273 is scaled down by a factor of 1800.}
\end{figure}

\begin{figure}[hbt]
\includegraphics[width=10.5cm, viewport=25 5 495 345]{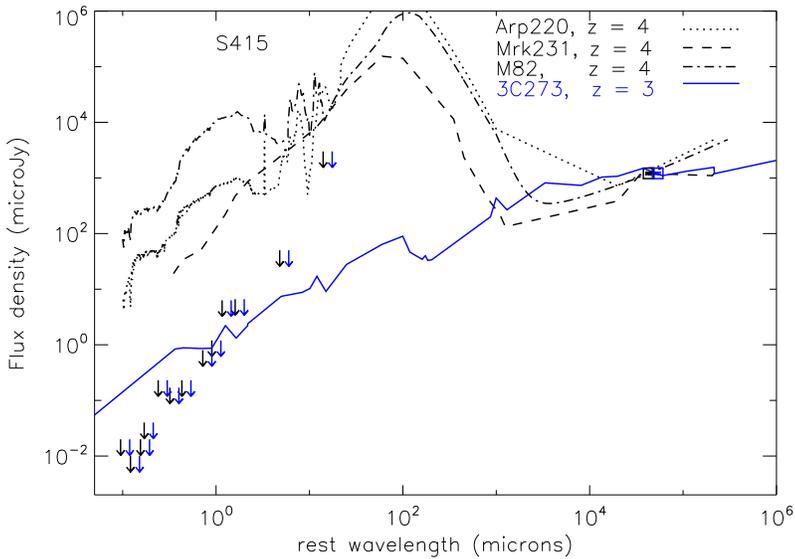}
\caption{The SED of S415, fit by Arp220, Mrk231, M82 and
3C273. Points in black assume the galaxy is at $z = 4$, and blue are the data for $z = 3$. This galaxy is not well fit
by M82, Arp220 or Mrk231, as it would be detected by {\sl Spitzer} imaging. The SED of radio-loud galaxy 3C273 at $z >> 1$ is
consistent with the observations, if obscuration at optical wavelengths is assumed. The SEDs are scaled up in luminosity by 2.7 $\times$ 10$^6$,  3.9 $\times$ 10$^5$ and 6.0  $\times$ 10$^7$ times for Arp220, Mrk231 and M82 respectively, and 3C273 is scaled down by a factor of 30.}
\end{figure}

\begin{figure*}[hbt]
\includegraphics[width=8.5cm, viewport=25 5 495 345]{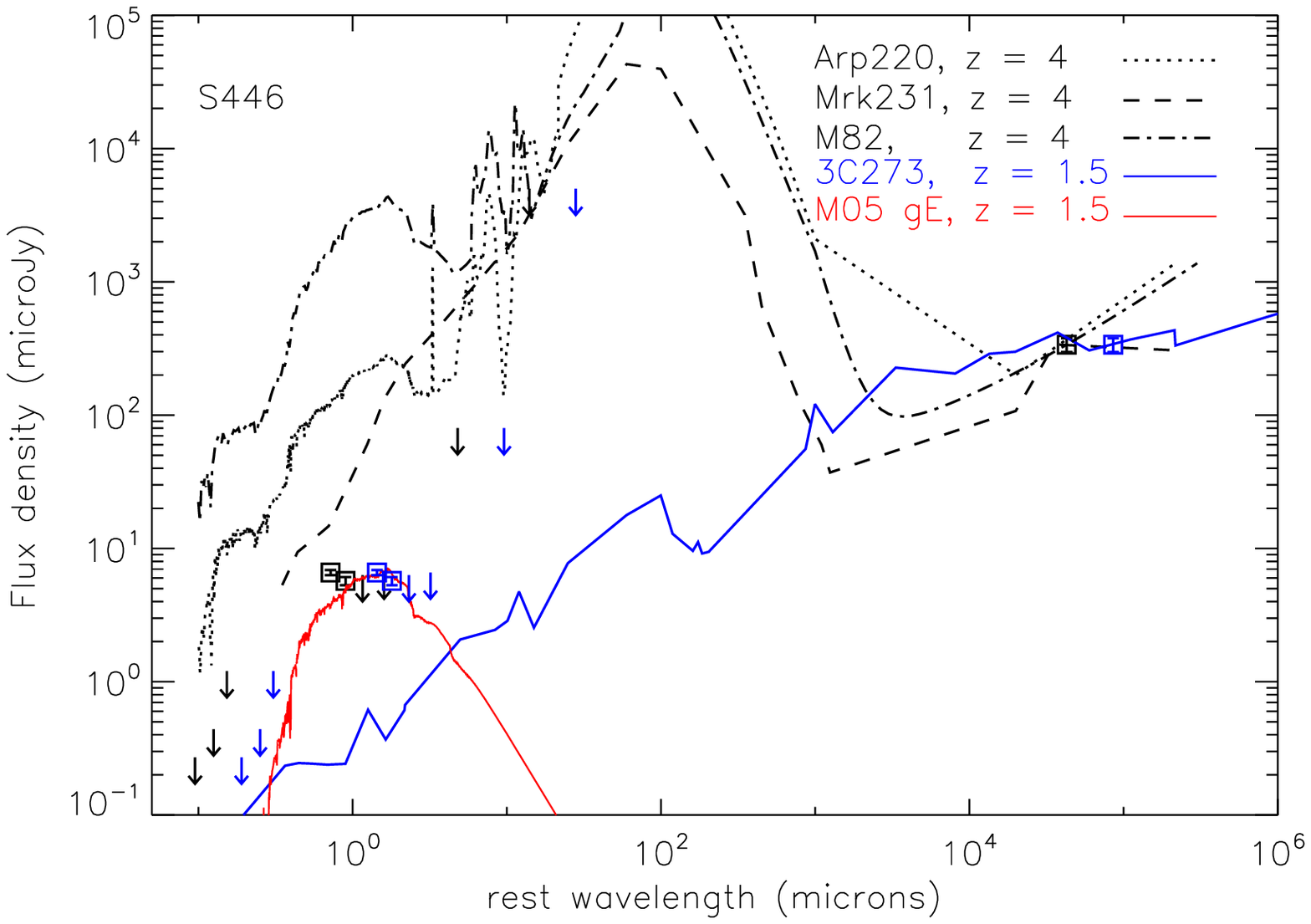}
\includegraphics[width=8.5cm, viewport=25 5 495 345]{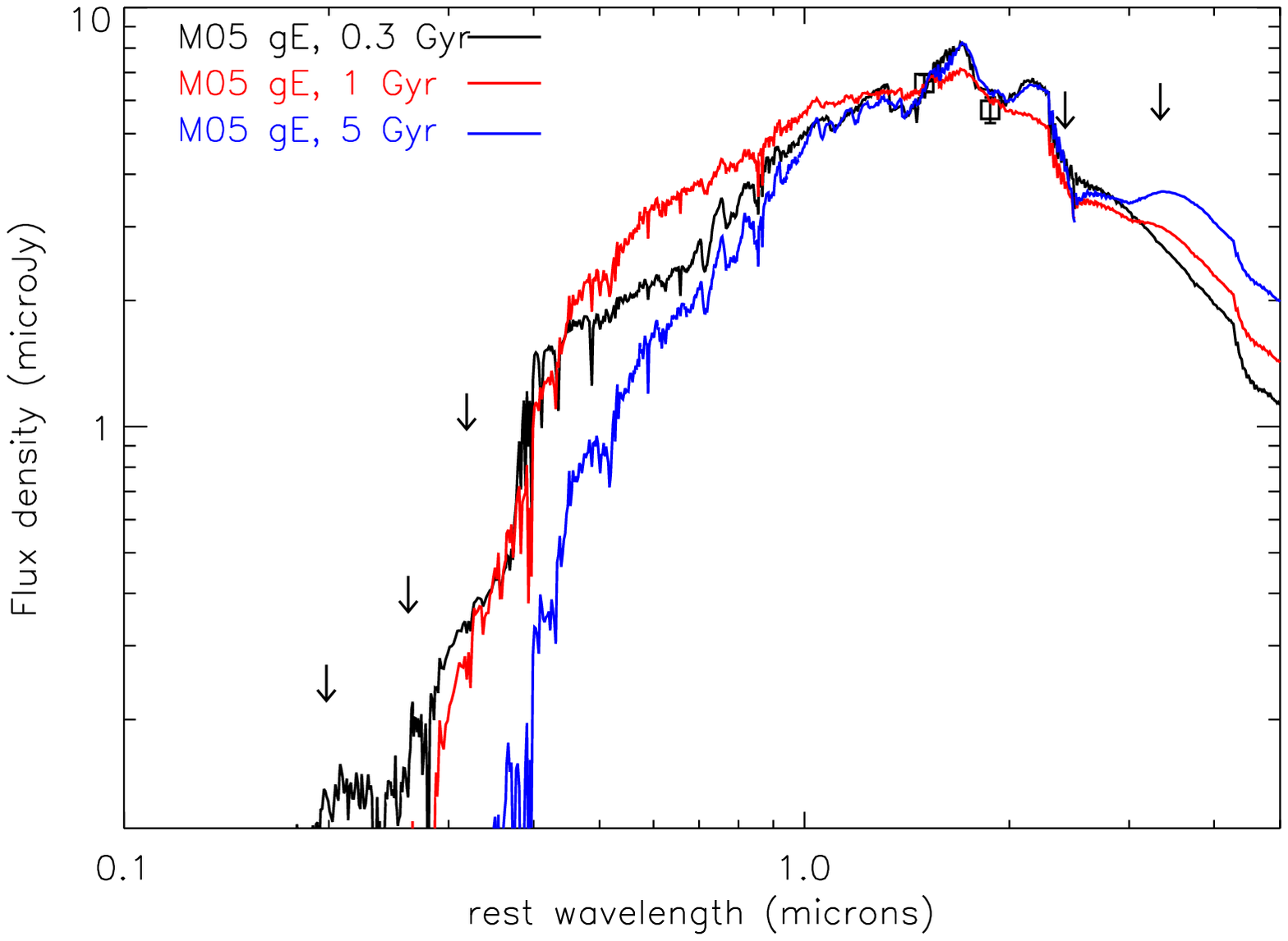}
\caption{Left: The SED of S446, fit by Arp220 Mrk231, M82 and 3C273. The SEDs are scaled up in luminosity by 750, 110 and 17000 times for Arp220, Mrk231 and M82 respectively, and 3C273 is scaled down by a factor of 550. An old stellar population (red line) from  \cite{maraston2005} is fit to the IRAC data. Points in black assume the
galaxy is at $z = 4$, and blue are the data for $z = 1.5$. This galaxy is not well fit by M82, Arp220 or Mrk231, as it would be
detected by 24 $\mu$m imaging. The SED of radio loud galaxy 3C273 at z $\sim$2.0 combined with the old stellar
population is consistent with the observations. Right: A zoom of the optical/NIR region. The \cite{maraston2005} stellar population models are shown at the best fit redshift of 1.5. The 0.3 and 5 Gyr models are satisfactory fits, but the 1 Gyr model provides the best fit. The optical limits are $g$, $r$ and $i$ limits from SWIRE. }
\end{figure*}

\begin{figure*}[hbt]
\includegraphics[width=8.5cm, viewport=25 5 495 345]{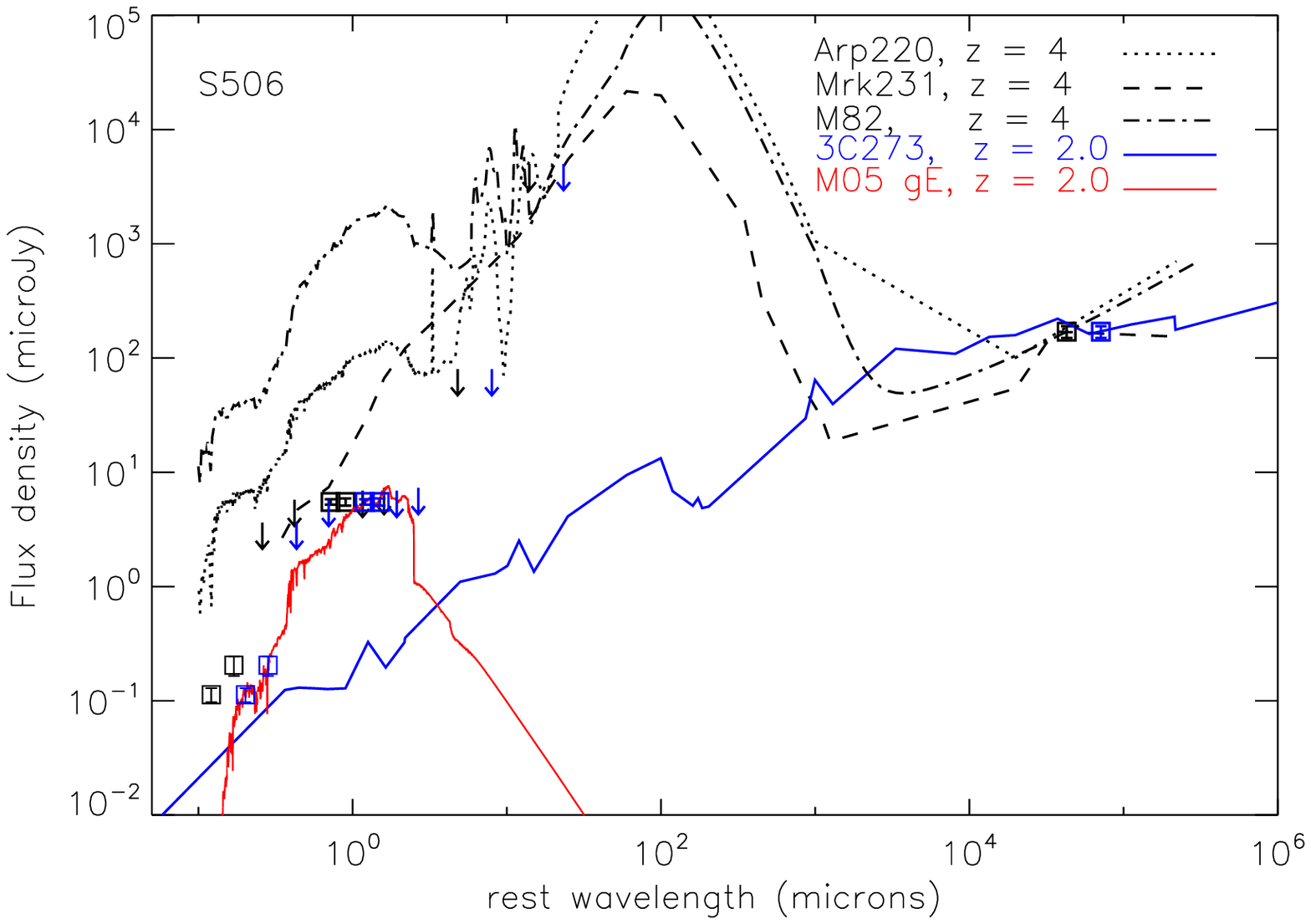}
\includegraphics[width=8.5cm, viewport=25 5 495 345]{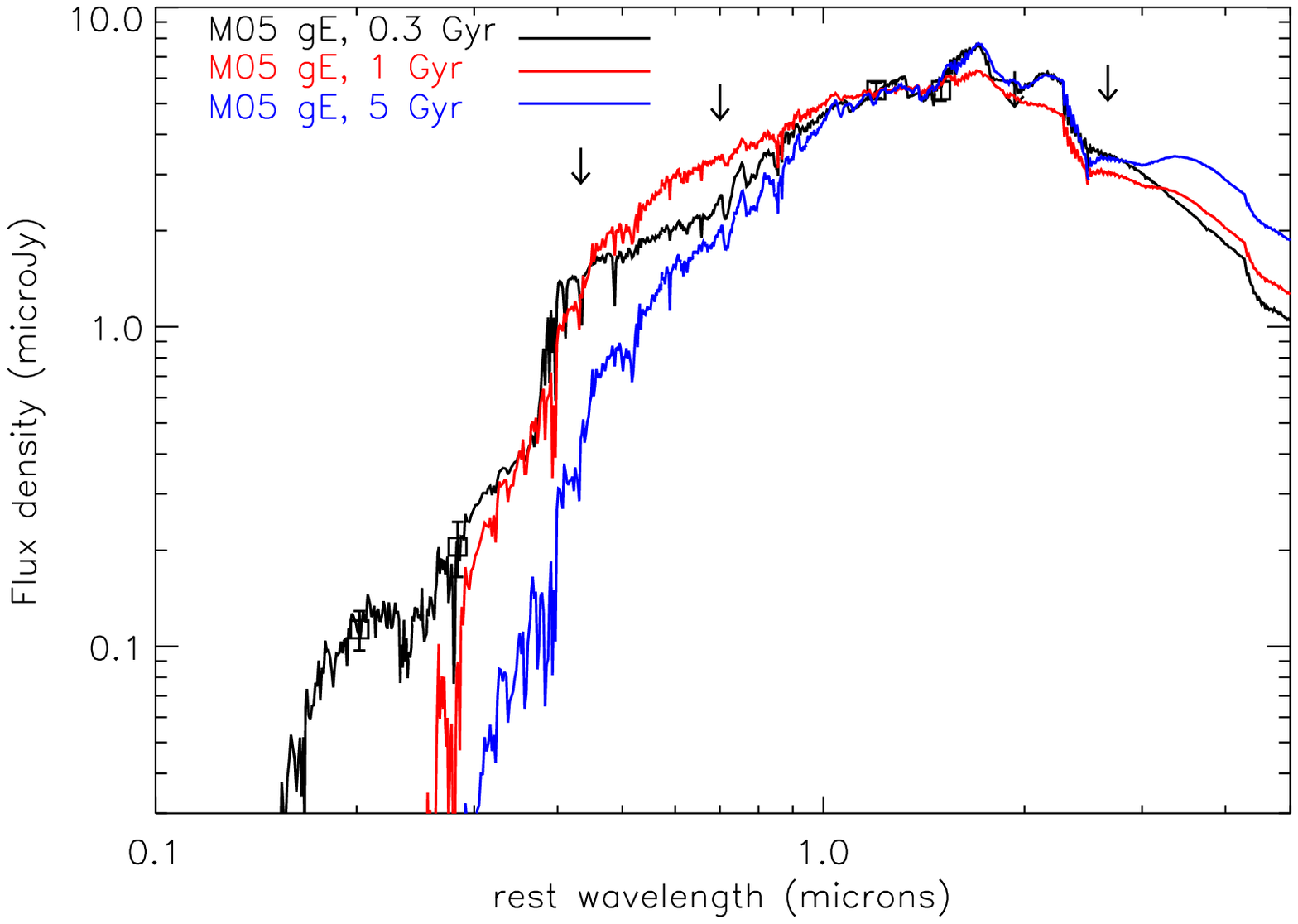}
\caption{Left: The SED of S506, fit by Arp220, Mrk231, M82 and 3C273. The SEDs are scaled up in luminosity by 380, 55 and 8500 times for Arp220, Mrk231 and M82 respectively, and 3C273 is scaled down by a factor of 530. An old stellar population (red line) from  \cite{maraston2005} 
is fit to the IRAC data. Points in black assume the galaxy is at z = 4, and blue are the data for $z = 2.0$. This galaxy is not well fit by M82, Arp220 or Mrk231, as it would be detected by 24 $\mu$m imaging. The SED of radio loud galaxy 3C273 at $z \sim 2.0$ combined with the old stellar population is consistent with the observations. Right: A zoom of the optical/NIR region. The \cite{maraston2005} stellar population models are shown at the best fit redshift of 2.0. The 1 and 5 Gyr models fit the IRAC data, but only the 0.3 Gyr model (black line) can reproduce both IRAC and ACS detections. The J and K limits are from MUSYC.}
\end{figure*}

\begin{figure}[hbt]
\includegraphics[width=8.5cm, viewport=25 5 485 340]{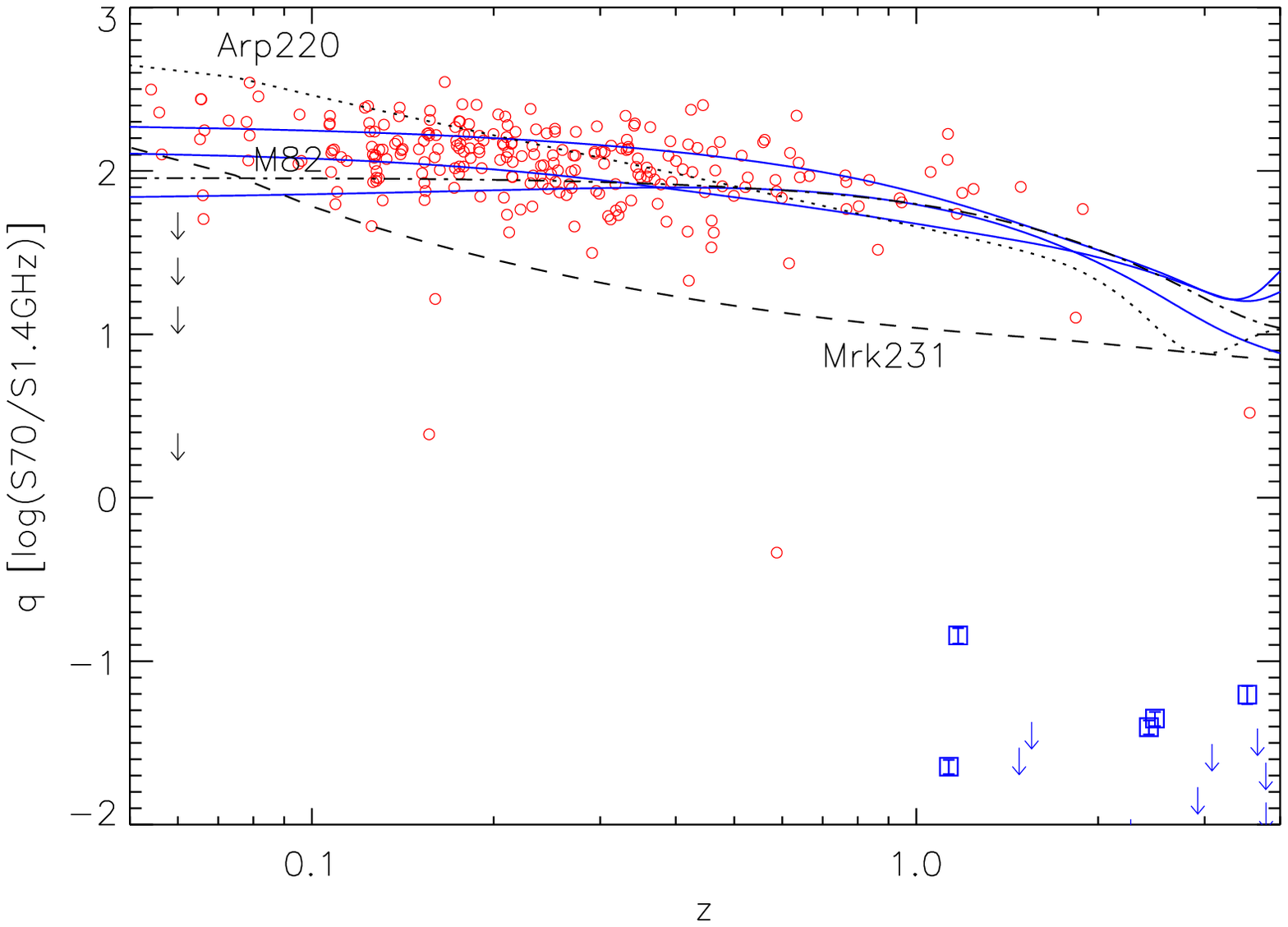}
\includegraphics[width=8.5cm, viewport=25 5 485 340]{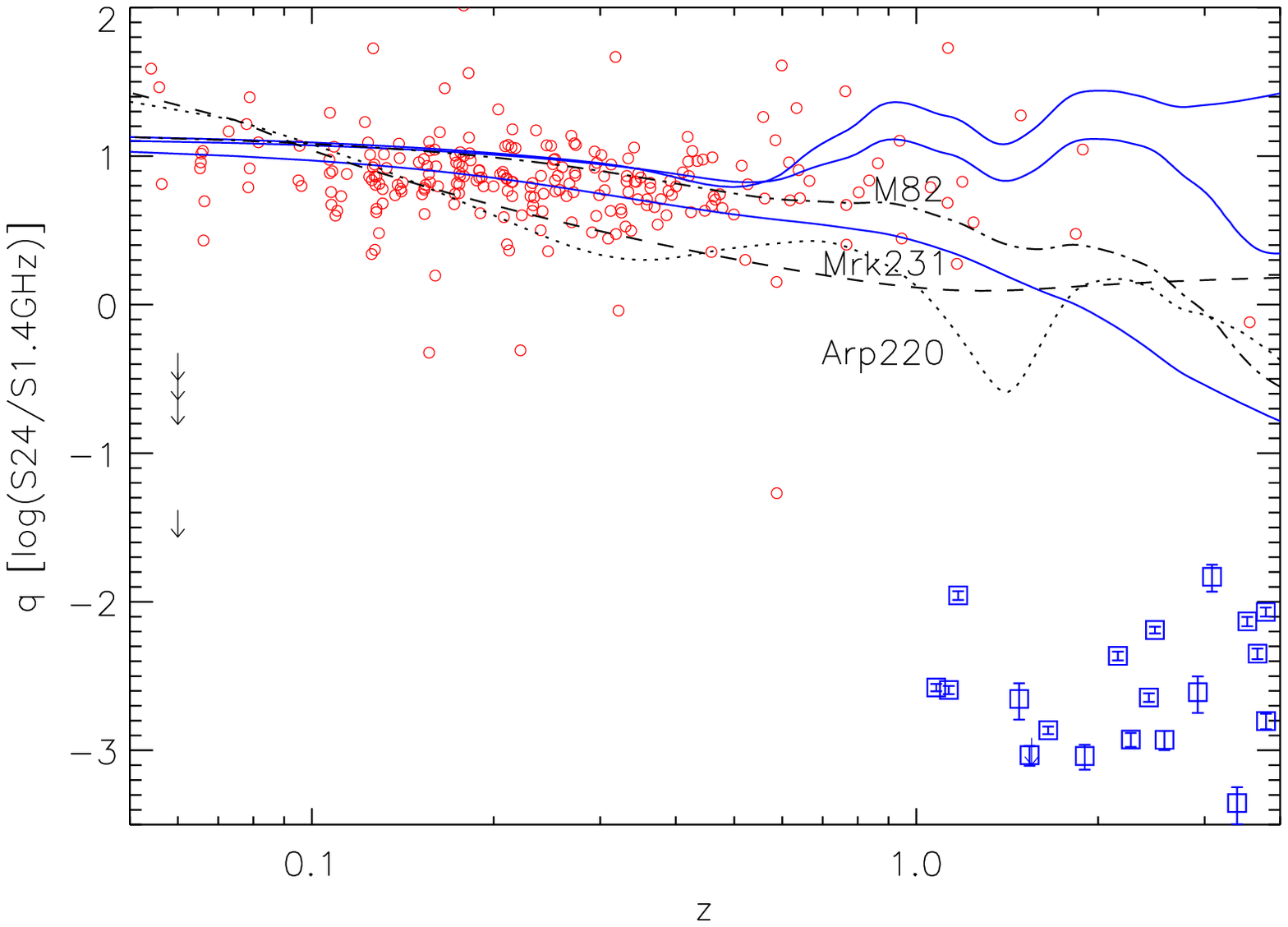}
\caption{The ratio of {\sl Spitzer} 70 (left) and 24 (right) $\mu$m to 1.4 GHz flux density, $q$, as a function of redshift. The four IFRSs are shown as black arrows at $z = 0.06$, as we have limits only. Red circles are xFLS galaxies \citep{frayer2006}, which are predominately star forming galaxies. Blue boxes or arrows show ratios from a sample of high redshift radio galaxies \citep{seymour07}, which are radio-loud AGN. The black SED tracks are Arp220 (dotted), Mrk231 (dashed) and M82 (dot dashed). Blue solid lines show tracks from \cite{ce01} with bolometric IR luminosities ranging from $10^8$ to $10^{13}$ $L_\odot$.}
\end{figure}

\begin{table*}[hbt]
\center
\begin{tabular}{lllll} \hline \hline
 & S283 &  S415 & S446 & S506 \\ \hline 
 RA (J2000) & 3:30:48.686 & 3:32:13.077 & 3:32:31.540 & 3:33:11.486 \\
 Dec (J2000) & -27:44:45.32 & -27:43:51.07 & -28:04:33.53 & -28:03:19.09\\
radio 1.4 GHz & 0.287 mJy & 1.21 mJy & 0.338 mJy & 0.170 mJy  \\ 
radio 2.4 GHz & $<0.40$ mJy & 0.67 mJy & $<0.45$ mJy & $<0.45$ mJy  \\ 
$\alpha^{2.5 {\rm GHz}}_{1.4 {\rm GHz}}$ ($S \propto \nu^\alpha$) & $<0.6$ & -1.1 $\pm$  0.13 & $<0.5$ & $<1.8$ \\ 
ACS B mag & \nodata & $>$28.1&  \nodata &  \nodata \\ 
ACS V mag &\nodata & $>$28.9  & \nodata &   26.27 \\ 
ACS I mag & \nodata & $>$28.3 & \nodata  &  \nodata \\ 
ACS z mag & \nodata & $>$27.4 &  \nodata &  25.62 \\ 
SWIRE g' mag & \nodata & - & $>$25.3 & - \\
SWIRE r' mag & \nodata & - & $>$24.8 & - \\
SWIRE i' mag & \nodata & - & $>$23.7 & - \\
J mag & \nodata & $>$25.5 &  \nodata  & $>$22.5 \\ 
H mag & \nodata & $>$25.8 &  \nodata  & \nodata \\ 
K mag & \nodata & $>$25.5 &  \nodata  & $>$22.0\\ 
IRAC 3.6 $\mu$m & $<$ 3.5 $\mu$Jy &  $<$ 0.8 $\mu$Jy &  6.6 $\pm$ 0.3 $\mu$Jy & 5.5 $\pm$ 0.3 $\mu$Jy \\ 
IRAC 4.5 $\mu$m & $<$ 7.0 $\mu$Jy & $<$ 1.2 $\mu$Jy & 5.7 $\pm$ 0.5 $\mu$Jy & 5.5 $\pm$ 0.4 $\mu$Jy \\ 
IRAC 5.8 $\mu$m & $<$ 41 $\mu$Jy & $<$ 6.3 $\mu$Jy & $<$ 6.3 $\mu$Jy & $<$ 6.3 $\mu$Jy  \\ 
IRAC 8.0 $\mu$m & $<$ 49 $\mu$Jy &  $<$ 6.6 $\mu$Jy & $<$ 6.6 $\mu$Jy & $<$ 6.6 $\mu$Jy \\ 
MIPS 24 $\mu$m & $<$ 100 $\mu$Jy & $<$ 50 $\mu$Jy & $<$ 80 $\mu$Jy & $<$ 80 $\mu$Jy \\ 
MIPS 70 $\mu$m &  $<$ 16 mJy & $<$ 3 mJy & $<$ 5 mJy &  $<$ 5 mJy \\  \hline
\end{tabular}
\caption{Summary of the observed properties of the four Infrared Faint Radio Sources in the extended Chandra Deep Field South.}
\end{table*}

\end{document}